\documentclass[conf]{new-aiaa}
\usepackage[utf8]{inputenc}

\usepackage{graphicx,wrapfig}
\graphicspath{{figures/}}
\usepackage{amsmath,booktabs}
\usepackage{bm}
\usepackage{xcolor}
\usepackage[version=4]{mhchem}
\usepackage{siunitx}
\usepackage{longtable,tabularx}
\usepackage{comment}
\usepackage{rotating}
\usepackage{subcaption}
\usepackage{savefnmark}

\newcommand{\reals}{{\mathbb{R}}}

\newcommand{\norm}[1]{\left\lVert#1\right\rVert}
\newcommand{\mnorm}[1]{{\left\vert\kern-0.25ex\left\vert\kern-0.25ex\left\vert #1 
    \right\vert\kern-0.25ex\right\vert\kern-0.25ex\right\vert}}

\newcommand{\mc}{\mathcal}

\title{MPC for momentum counter-balanced and zero-impulse contact with a free-spinning satellite}

\author{Theofania Karampela\footnote[1]{Both authors contributed equally.}}
\affil{ETH Z\"urich,  Z\"urich, Switzerland, 8092}

\author{Rishie Seshadri*}
\affil{Georgia Institute of Technology, Atlanta, Georgia, 30332}

\author{Florian D\"orfler}
\affil{ETH Z\"urich,  Z\"urich, Switzerland, 8092}

\author{Sarah H.Q. Li}
\affil{Georgia Institute of Technology, Atlanta, Georgia, 30332}
\begin{document}

\maketitle

\begin{abstract}
In on-orbit robotics, a servicer satellite's ability to make contact with a free-spinning target satellite is essential to completing most on-orbit servicing (OOS) tasks. This manuscript develops a nonlinear model predictive control (MPC) framework that generates feasible controls for a servicer satellite to achieve zero-impulse contact with a free-spinning target satellite. The overall maneuver requires coordination between two separately actuated modules of the servicer satellite: (1) a moment generation  module and (2) a manipulation module. We apply MPC to control both modules by explicitly modeling the cross-coupling dynamics between them. We demonstrate that the MPC controller can enforce actuation and state constraints that prior control approaches could not account for. We evaluate the performance of the MPC controller by simulating zero-impulse contact scenarios with a free-spinning target satellite via numerical Monte Carlo (MC) trials and comparing the simulation results with prior control approaches. Our simulation results validate the effectiveness of the MPC controller in maintaining spin synchronization and zero-impulse contact under operation constraints, moving contact location, and observation and actuation noise. 
\end{abstract}



\section{Introduction}
\label{sec:Intro}
In recent years, the low Earth orbit's (LEO) expanding satellite population has led to an emerging need for
OOS: the observation or alteration of satellites after initial launch by servicer satellites~\cite{duke2021orbit}.
Central to this effort is the development of servicer satellites that can autonomously execute essential OOS maneuvers including refueling, attitude stabilization, debris removal, and in-orbit assembly~\cite{Aghili2012,Lampariello2018}. Each of these tasks  require contact with orbital objects that typically lack coordination capabilities.

While analogous maneuvers are routinely performed by ground-based robotic systems using optimal control and well-characterized kinematic models~\cite{Zeilinger, Kara2023}, executing these same maneuvers in orbit is substantially more challenging.
Executing these maneuvers in orbit involves free-flying, momentum-coupled multibody systems whose tightly coupled translational and rotational rigid-body dynamics complicate the usage of optimal control frameworks~\cite{Space_Robotics,Aghili2024,Alizadeh2024,giordano2016dynamics}.
Most ground-based maneuvers of manipulation arms are performed with anchored arm bases, whereas in OOS, the manipulation arm base is freely floating, introducing complex rotational dynamics~\cite{giordano2016dynamics}. A key challenge in achieving zero-impulse contact is the momentum coupling between the manipulation module and the moment-generation module: motion in the manipulation module exerts reaction torques on the moment-generation module, which affects the servicer satellite's overall orientation. In this case, the moment-generation module's reaction wheel (RW) cluster must counteract the reaction torques from the arm to keep the servicer satellite steady~\cite{Space_Robotics,Aghili2024}.

In this manuscript, we develop autonomous control approaches for a momentum-controlled servicer satellite equipped with a moment-generation module and a manipulation module to establish contact with a free-spinning target satellite. We decompose this operation into two consecutive phases~\cite{Aghili2024}:
\begin{enumerate}
    \item Spin and orientation synchronization with the target satellite;
    \item Zero-impulse contact at a designated location with the target satellite. 
\end{enumerate} 

Prior control approaches~\cite{Aghili2024} risk violating the servicer satellite's actuation constraints, which can lead to critical system failures for the actuation modules. Additionally, actuation constraints on the servicer satellite's moment-generation module couple the servicer satellite's rotational dynamics to the manipulation module's forward kinematics. In our approach, we explicitly capture actuation limits of the actuation modules by utilizing a nonlinear MPC controller~\cite{Zeilinger}.


\textbf{Contributions.}
We propose a nonlinear MPC control framework to generate optimal control that enables a servicer satellite to make zero-impulse contact with a free-spinning target satellite under operational constraints. 
To enable MPC usage, we explicitly derive the servicer satellite's state-based and rigid body dynamics using quaternions, satellite positions, satellite angular velocities, and arm joint angles. We then derive a reduced state formulation to enable faster MPC computation and incorporate the dynamics of the manipulation arm's end effector via forward kinematics. We solve the resulting MPC using a publicly available  MPC solver \texttt{acados}~\cite{acados}, and evaluate its numerical speed on simulated contact scenarios. We then present a comparative analysis of our MPC control approach against prior control approaches to evaluate performance across various scenarios and demonstrate the key advantages of MPC. 

\subsection{Literature review}
\label{sec:Lit_Review}
Interest in OOS has grown rapidly in recent years, driving many space applications in this direction~\cite{FLORESABAD20141, Alizadeh2024, Freeman2020}.
OOS is a broad domain that spans repair, refueling, component upgrades, inspection, and debris removal~\cite{papadopoulos2021robotic, Zhang2022, ODQN}.
Executing OOS tasks requires the servicer satellite to autonomously stabilize itself under actuation limits, environmental constraints, and real-time disturbances~\cite{Alizadeh2024}. For example, solar-powered satellites must preserve precise panel and antenna orientations under disturbances like atmospheric drag~\cite{Starin2005, Colagrossi2022}.
A key subtask in OOS is detumbling, i.e. stabilizing a tumbling target by reducing its  angular velocity~\cite{Aghili2009_Time_Optimal_Detumbling, INVERNIZZI2020108779}. 
Detumbling restores orientation control to satellites and can significantly reduce the risk of future collisions~\cite{Zhang2022, Bennett2025}.
Prior control approaches to detumbling  include PID and Lyapunov-based controllers~\cite{Aghili2024, gao2024adaptive} and optimal control methods that explicitly minimize fuel consumption and/or detumbling time~\cite{Aghili2012, Aghili2009, Aghili2019}.

 MPC supports a wide range of real-time control in industry, such as robotic manipulation~\cite{FORBES2015531, QIN2003733}, automotive systems~\cite{Hrovat2012, electronics10212593}, and aerospace guidance and attitude control~\cite{Cairano2012ModelPC, Guiggiani}.
MPC enables high-precision trajectory tracking on multi-degrees of freedom (DoF) arms under model uncertainty and constraints~\cite{Zeilinger}, as well as on low-DoF manipulators where lightweight setups still require strict constraint handling~\cite{Kara2023, inproceedings}.
Nonlinear MPC further enforces safety constraints during the manipulation and motion planning of complex dynamical systems~\cite{Tereda2023, 7984201, BuizzaAvanzini2018}. When used to station-keep for in-orbit satellites, MPC demonstrates strong capabilities in constraint handling and momentum management by regulating attitude and controlling momentum while enforcing actuation and safety limits~\cite{Guiggiani, 8062723}.
\section{Servicer satellite dynamics modeling}
In this section, we outline the servicer satellite's dynamics and kinematics models, which elaborates the model from~\cite{Aghili2024}. We assume that  the servicer satellite is composed of rigid bodies and has known inertia properties. We derive reduced state representations that we integrate into the MPC solver in Section~\ref{sec:MPC_framework}. We further assume that the servicer knows the spin orientation and rate of the target satellite. This assumption is valid when the servicer satellite has on-board sensors and estimation algorithms~\cite{liu2021guidance}. Finally, we assume the environment exerts no external forces on either the servicer or the target satellite, and that there is zero relative linear velocity between the centroids of the servicer and target satellites.

The servicer satellite primarily consists of two independent actuation modules: a moment-generation module and a manipulation module. We model the moment-generation module as a cuboid base with a $3$-DoF momentum-balancing RW cluster aligned with the principal axes of the moment-generation base, and the manipulation module as a $3$-joint, $3$-DoF rigid body arm. These models simplify realistic servicer satellites but are representative of the servicer satellite's actuation modules that are relevant to making contact with a free-spinning target satellite. 

\textbf{Notation}. We use  $[X]$ to denote the set of natural numbers between $1$ and $X$ inclusive, $[a,b]$ to denote the set of natural numbers between $a$ and $b$ inclusive,  $\reals^n$  to denote the real number vector space with dimension $n$, and $\bm{v}$ to denote vectors whose dimension is greater than one. 

\subsection{Moment-generation module: 3-DoF RW cluster and cuboid base}
The moment-generation module serves as the primary actuation system for controlling the servicer satellite's attitude and angular velocity. 
It consists of a cuboid base equipped with a 3-DoF RW cluster comprising three identical actuators~\cite{GIORDANO}. We model each RW to align with one of the three principal axes of the servicer satellite's cuboid base. This configuration enables independent torques along each body axis, where the total torque generated is 
\begin{wrapfigure}{r}{0.52\textwidth}
\vspace{-2pt}
  \begin{center}
    \includegraphics[width=.51\textwidth]{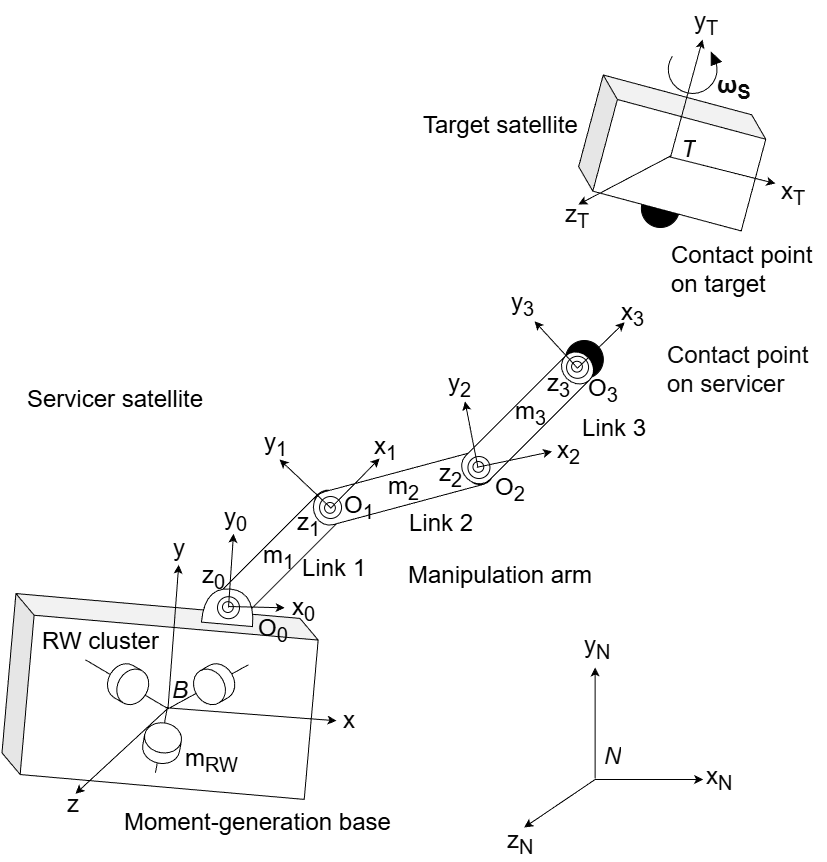}
  \end{center}
  \caption{Servicer satellite with a $3$-DoF manipulation arm and a moment-generation base.}\label{fig:space_robot_schematic}
\end{wrapfigure}
\begin{equation}
    \label{eq:rw_torque}
    \bm{\tau_r} = \begin{bmatrix} \tau_{r,x} & \tau_{r,y} & \tau_{r,z} \end{bmatrix} \in \reals^3,
\end{equation}
and the total angular displacement of the three RWs as 
\begin{equation}
    \bm{\phi} = \begin{bmatrix}
    \phi_x & \phi_y & \phi_z
    \end{bmatrix} \in \reals^3.
\end{equation}
The moment-generation module aims to align the servicer satellite's spin orientation and synchronize its angular velocity with respect to the target satellite. We denote the servicer satellite base's angular velocity as $\bm{\omega_B} \in \reals^3$, expressed in the body-fixed frame $B$ located at the center of mass (CoM) of the moment-generation base as shown in Figure~\ref{fig:space_robot_schematic}. Similarly, we denote the angular velocity of the target satellite as $\bm{\omega_S} \in \reals^3$, expressed in the target satellite's body-fixed frame $T$ located at the target satellite's CoM. We also introduce an inertial orbital frame $N$, shown in Figure~\ref{fig:space_robot_schematic}, which serves as the global reference frame with respect to which the attitudes of both frames $B$ and $T$ are defined.

We use quaternions, which provide a singularity-free representation for three-dimensional orientations, to represent the servicer satellite's relative orientation to the target satellite~\cite{Aghili2012}. A quaternion is a column vector $\bm{q} = \begin{bmatrix} \bm{q_{\upsilon}}^{\top} & q_{0} \end{bmatrix} \in \reals^4$, where $\bm{q_{\upsilon}} \in \reals^3$ denotes the {vector component} and $q_{0} \in \reals$ is the {scalar component}~\cite{voight2021quaternion}. We use $D_{\bm{q}} = \left\{ \bm{q} \in \reals^4 \mid \|\bm{q}\|_2 = 1 \right\}$ to denote the set of all unit quaternions, where $\|\bm{q}\|_2 = \sqrt{\bm{q}^\top \bm{q}} \in \reals$ is the Euclidean norm. In the following sections, we utilize the quaternion operators $\otimes$, $\star$, $-1$, respectively defined as 
\begin{equation}
\bm{a} \otimes \bm{b} :=
    \begin{bmatrix}
    a_0 \bm{b_{\upsilon}} + b_0 \bm{a_{\upsilon}} + \bm{a_{\upsilon}} \times \bm{b_{\upsilon}} \\
    a_0 b_0 - \bm{a_{\upsilon}}^\top \bm{b_{\upsilon}}
    \end{bmatrix} \in \reals^4, \quad \bm{a}^\star := \begin{bmatrix} -\bm{a_{\upsilon}}^\top & a_0 \end{bmatrix} \in \reals^4, \quad \bm{a}^{-1} := \frac{\bm{a}^\star}{\|\bm{a}\|_2^2} \in \reals^4, \quad \forall \bm{a}, \bm{b} \in D_{\bm{q}}. 
\end{equation}

Let $\bm{q}_B \in D_{\bm{q}}$ denote  the servicer satellite base frame $B$'s orientation, and let $\bm{q}_T \in D_{\bm{q}}$ denote the target satellite frame $T$'s orientation. The relative orientation quaternion from frame $T$ to frame $B$ is given by
\begin{equation} \label{eq:rel_quat}
    \bm{q_{\textbf{rel}}} := \bm{q}_B \otimes \bm{q}_T^{-1} =   \begin{bmatrix}
        \bm{q_{\upsilon,\textbf{rel}}}^\top & {q}_{0,\text{rel}}
    \end{bmatrix} \in D_{\bm{q}},\quad \bm{q_{\upsilon,\textbf{rel}}} \in \mathbb{R}^3,\; {q}_{0,\text{rel}} \in \mathbb{R}.
\end{equation}
To ensure that the relative orientation quaternion is represented as a pure rotation and to prevent numerical errors from accumulating, we require  $\bm{q_{\textbf{rel}}} \in D_{\bm{q}}$, i.e., the relative orientation quaternion must remain normalized. 
We compute the relative angular velocity of the servicer satellite in  frame $B$ using the relative orientation quaternion $\bm{q_{\textbf{rel}}}$. We first define a quaternion rotation matrix $A(\bm{q_{\textbf{rel}}}) = I_{3\times3} + 2 q_{0,\text{rel}}[\bm{q_{\upsilon,\textbf{rel}}}]_\times + 2[\bm{q_{\upsilon,\textbf{rel}}}]_\times^2 \in \reals^{3\times3}$, which transforms vectors from the target frame $T$ to the servicer frame $B$,
where $[\bm{q_{\upsilon,\textbf{rel}}}]_\times \in \reals^{3\times3}$ is the skew-symmetric matrix of $\bm{q_{\upsilon,\textbf{rel}}}$~\cite{voight2021quaternion}. 
The relative angular velocity $\bm{\omega_{\textbf{rel}}} \in \reals^3$ can then be expressed as
\begin{equation}
\label{eq:omega_rel}
\bm{\omega_{\textbf{rel}}} = \bm{\omega_B} - A(\bm{q_{\textbf{rel}}}) \bm{\omega_S} \in \reals^3,
\end{equation}
where $A(\bm{q_{\textbf{rel}}})$ transforms the target's angular velocity $\bm{\omega_S}$ into the servicer satellite's body frame $B$. 
Finally, the relative orientation quaternion $\bm{q_{\textbf{rel}}}$ evolves over time based on the relative angular velocity via the quaternion dynamics~\cite{Aghili2024} given by
\begin{equation}
\label{eq:quaternion_kinematics}
{\bm{\dot{q}}_{\textbf{rel}}}(t) = \frac{1}{2}
\begin{bmatrix}
-[\bm{\omega_{\textbf{rel}}}]_\times & \bm{\omega_{\textbf{rel}}} \\
-\bm{\omega_{\textbf{rel}}}^\top & 0
\end{bmatrix}_{4\times4}
\bm{q_{\textbf{rel}}}(t) \in \reals^4. 
\end{equation}
Together, equations~\eqref{eq:omega_rel} and~\eqref{eq:quaternion_kinematics} capture the servicer satellite base dynamics, represented by the state variables $\bm{q_{\textbf{rel}}}$ and $\bm{\omega_{\textbf{rel}}}$, and controlled by the control torque $\bm{\tau_r}$.

\subsection{Manipulation module: 3-DoF manipulation arm}
The manipulation module serves as the primary actuation system for achieving zero-impulse contact with the target satellite. It consists of a $3$-DoF manipulation arm with three revolute joints and an end effector. 
We denote the module's configuration  by each joint's angular displacement $\bm{\theta} = \textstyle\begin{bmatrix}
    \theta_1 & \theta_2 & \theta_3
\end{bmatrix} \in [0,2\pi]^3$ and each joint's  angular velocity $\bm{\dot{\theta}} = \begin{bmatrix}
    \dot{\theta}_1 & \dot{\theta}_2 & \dot{\theta}_3
\end{bmatrix} \in [0,2\pi]^3$. Each joint is also equipped with joint actuators that apply rotational torques, denoted as 
\begin{equation}\label{eqn:joint_torques}
    \bm{\tau_m} = \begin{bmatrix}
    \tau_{m,1} & \tau_{m,2} & \tau_{m,3}
    \end{bmatrix} \in \reals^3.
\end{equation} 
We denote the end effector position as $\bm{p_{ee}} \in \reals^3$ and the end effector velocity as $\bm{v_{ee}} \in \reals^3$, expressed in the servicer satellite base frame $B$ (Figure~\ref{fig:space_robot_schematic}). To link the end effector position to the joint configuration, we derive the forward kinematics using the Denavit-Hartenberg (DH) convention~\cite{denavit1955kinematic, Siciliano2010Robotics}, which systematically relates the end effector pose to the joint angles through homogeneous transformation matrices.
As shown in Figure~\ref{fig:space_robot_schematic}, each link $i$ in our $3$-DoF manipulation arm is characterized by four DH parameters: link length $L_i$, link twist $\alpha_i$, link offset $d_i$, and joint angle $\theta_i$. We assume that  all joints are revolute, all offsets $d_i = 0$ and all link twists $\alpha_i = 0$. 
For simplicity, we consider a planar 3-DoF manipulator operating in the $xy$ plane such that $z_{ee}=\dot{z}_{ee}=0$.
For this manipulator configuration with link lengths $L_1, \, L_2, \, L_3$, the forward kinematics are explicitly derived as 
\begin{equation} \label{eq:fk_position}
    \bm{p_{ee}} = \begin{bmatrix}
        x_{ee} \\
        y_{ee} \\
        z_{ee}
    \end{bmatrix} = \begin{bmatrix}
        L_1 \cos{\theta_1} + L_2 \cos{(\theta_1 + \theta_2)} + L_3 \cos{\theta_{ee}} \\
        L_1 \sin{\theta_1} + L_2 \sin{(\theta_1 + \theta_2)} + L_3 \sin{\theta_{ee}} \\
        0
    \end{bmatrix} \in \reals^3, \quad
    \theta_{ee} = \theta_1 + \theta_2 + \theta_3 \in \reals. 
\end{equation}
Similarly, the end effector velocity $\bm{v_{ee}} \in \reals^3$ is given by 
\begin{equation} \label{eq:fk_velocity}
    \bm{v_{ee}} = \begin{bmatrix}
        \dot{x}_{ee} \\
        \dot{y}_{ee} \\
        \dot{z}_{ee}
    \end{bmatrix}= \begin{bmatrix}
        -L_1 \sin{\theta_1} \dot{\theta}_1 - L_2 \sin{(\theta_1 + \theta_2)}(\dot{\theta}_1 + \dot{\theta}_2) - L_3 \sin{\theta_{ee}}(\dot{\theta}_{ee}) \\
        L_1 \cos{\theta_1} \dot{\theta}_1 + L_2 \cos{(\theta_1 + \theta_2)}(\dot{\theta}_1 + \dot{\theta}_2) + L_3 \cos{\theta_{ee}}(\dot{\theta}_{ee}) \\
        0
    \end{bmatrix}.
\end{equation} 
The forward kinematics enable the joint torques $\bm{\tau_m}$~\eqref{eqn:joint_torques} to manipulate the end effector position $\bm{p_{ee}} $ and velocity $ \bm{v_{ee}}$.  

\subsection{Momentum-coupled dynamics derivation}
\label{sec:moment_couple_dynamics}
One of the key challenges in controlling the servicer satellites is accounting for the coupled momentum between the cuboid base ($\bm{q_{\textbf{rel}}}, \, \bm{\omega_{\textbf{rel}}}$), the RW cluster ($\bm{\phi}$) and the manipulation arm ($\bm{\theta}$, $\bm{v_{ee}}$, $\bm{p_{ee}}$). In this section, we leverage the law of momentum conservation to derive the servicer satellite's state dynamics. These derivations were first introduced and more succinctly presented in~\cite{Aghili2024}.

Based on~\cite{Space_Robotics}, we can derive the inertia matrix for a free-flying multi-body system with moment-generation base.
Consider the multi-body system in Figure~\ref{fig:space_robot_schematic}, we index rigid bodies of the servicer satellite by $i \in [0,6]$, where $i=0$ denotes the cuboid base, $i \in [1,3]$ denote the three links of the manipulator arm, and $i \in [4,6]$ denote the three RWs of the RW cluster. The cuboid base ($i=0$) has linear and angular velocities $\bm{v_B}, \bm{\omega_B} \in \reals^3$ in frame $B$, both of which are known. For each component $i \in [1,6]$, $\bm{v_i}, \, \bm{\omega_i} \in \reals^3$ denote the linear and angular velocities in frame $B$, while $\bm{\hat{v}_i}, \, \bm{\hat{\omega}_i} \in \reals^3$ denote the same velocities in the orbital inertial frame $N$.

If the manipulation arm has known joint angular velocity $\dot{\bm{\theta}} \in \reals^3$, then  each arm link's translation velocity and angular velocities in their respective body inertia frames are given  by 
\begin{equation} \label{eq:arm_velocities}
    \textstyle \bm{v_i} = J_{L_i}(\bm{\theta}) \bm{\dot{\theta}} \in \reals^{3}, \quad
\bm{\omega_i} = J_{A_i}(\bm{\theta}) \bm{\dot{\theta}} \in \reals^{3}, \forall \ i \in [1,3],
\end{equation}
where  $J_{L_i}(\bm{\theta}), \, J_{A_i}(\bm{\theta}) \in \reals^{3\times 3}$ are each arm link's joint angle-dependent and  geometry-dependent linear and angular Jacobian matrices respectively~\cite{Spong2006}.
For the RW cluster, we assume purely rotational motion about the $B$ frame's axes and no translational motion, thus
\begin{equation} \label{eq:RW_velocities}
    \textstyle \bm{v_i} = \bm{0}_{3 \times 1}, \quad
\bm{\omega_i} = J_{A_i}(\bm{\phi}) \bm{\dot{\phi}} \in \reals^{3}, \forall \ i \in [4,6].
\end{equation}
Let $\bm{r}_i$ denote the position of body $i$'s CoM relative to frame $B$. The translational velocity $\bm{\hat{v}_i}$ and angular velocity $\bm{\hat{\omega}_i}$ of body $i$ in the orbital inertial frame $N$ can be derived as $\textstyle \bm{\hat{v}_i} = \bm{v_i} + \bm{v_B} + \bm{\omega_B} \times \bm{r_i} \in \reals^{3}, \,
\bm{\hat{\omega}_i} = \bm{\omega_i} + \bm{\omega_B} \in \reals^{3}, \, \forall i \in [1,6].$
From these velocity terms and the known mass and inertia parameters $m_0 \in \reals_+, \, I_0 \in \reals^{3\times 3}$ for the moment-generation base, and $m_i \in \reals_+, \ I_i \in \reals^{3\times 3}$ for $i \in [1,6]$, the servicer satellite's total kinetic energy in the orbital inertial frame $N$ can be expressed as $  K = \frac{1}{2} \sum_{i=0}^{6} \big(\bm{\hat{v}_i}^\top m_i \bm{\hat{v}_i} + \bm{\hat{\omega}_i}^\top I_i \bm{\hat{\omega}_i} \big) \in \reals_+.$
We can substitute in $\bm{\hat{v}_i}, \bm{\hat{\omega}_i}$ to derive that the total kinetic energy $K$ is a quadratic function of the satellite states $    \textstyle   \bm{w} =  \begin{bmatrix} \bm{v_B} & \bm{\omega_B} & \bm{\dot{\theta}} & \bm{\dot{\phi}}\end{bmatrix} \in \reals^{12}$ given by 
\begin{equation} \label{eq:kinetic_energy}
 \textstyle K( \bm{v_B} ,\bm{\omega_B},\bm{\dot{\theta}}, \bm{\dot{\phi}}) = \bm{w}^\top H \bm{w} = \begin{bmatrix}
        \bm{v_B}^\top & \bm{\omega_B}^\top  & \bm{\dot{\theta}}^\top  & \bm{\dot{\phi}}^\top 
    \end{bmatrix}
    \begin{bmatrix}
        H_{\textbf{V}} & H_{\textbf{V} {\bm{\Omega}}} & H_{\textbf{V} {\bm{\theta}}} & H_{\textbf{V}\bm{\phi}} \\
        H_{\textbf{V} {\bm{\Omega}}}^\top & H_{{\bm{\Omega}}} & H_{{\bm{\Omega}} {\bm{\theta}}} & H_{{\bm{\Omega}} \bm{\phi}} \\
        H_{\textbf{V} {\bm{\theta}}}^\top & H_{{\bm{\Omega}} {\bm{\theta}}}^\top & H_{{\bm{\theta}}} & H_{{\bm{\theta}} \bm{\phi}} \\
        H_{\textbf{V}\bm{\phi}}^\top & H_{{\bm{\Omega}} \bm{\phi}}^\top & H_{{\bm{\theta}} \bm{\phi}}^\top & H_{\bm{\phi}}
    \end{bmatrix}
    \begin{bmatrix}
        \bm{v_B} \\ \bm{\omega_B} \\ \bm{\dot{\theta}} \\ \bm{\dot{\phi}}
    \end{bmatrix}, 
\end{equation}
where the block matrices $H_{ij}$ for $i, j \in \{\textbf{V}, \, {\bm{\Omega}}, \, {\bm{\theta}}, \, \bm{\phi}\}$ are provided in Appendix~\ref{APP:inertia_deriv}. The $H$ matrices are  the inertia matrices of the multi-body system, also defined in~\cite{Space_Robotics}, and are different from the generalized inertia matrices $M$ from~\cite{Aghili2024}.

Next, we derive the servicer satellite's system dynamics using the Euler-Lagrange equation as well as the generalized inertia matrices $M$ and the generalized coriolis and centrifugal forces $\bm{c}$. We define the servicer satellite's Lagrangian as the difference between its total kinetic energy $K$ and its total potential energy~\cite{Space_Robotics, Spong2006}. 
Since the relative potential energy is zero within the same orbit, the Euler-Lagrange equation simplifies to
\begin{equation} \label{eq:Euler_Lagrange_simp}
\frac{d}{dt} \left( \frac{\partial K}{\partial \bm{\dot{w}}} \right) - \frac{\partial K}{\partial \bm{w}} = \bm{\tau},
\end{equation}
where \( K = K(\bm{w}, \bm{\dot{w}}, t) \) is the total kinetic energy of the system, \( \bm{w} \in \reals^{12} \) represents the generalized coordinates, \( \bm{\dot{w}} = \frac{d\bm{w}}{dt} \in \reals^{12}\) is the time derivatives of the generalized coordinates, and $\bm{\tau} = \begin{bmatrix}
        \bm{f_{bV}}& \bm{f_{b\Omega}} & \bm{\tau_m} & \bm{\tau_r}
    \end{bmatrix}  \in \reals^{12}$ denotes the vector of generalized forces: 
    $\bm{f_{bV}} \in \reals^3$ is the linear force applied at the moment-generation base of the servicer satellite,
    $\bm{f_{b\Omega}} \in \reals^3$ is the torque applied to the moment-generation base,
    $\bm{\tau_m} \in \reals^3$ are the joint torques of the manipulation arm, and
    $\bm{\tau_r} \in \reals^3$ are the RW cluster torques. Expanding equation~\eqref{eq:Euler_Lagrange_simp} gives us the full servicer satellite dynamics as follows
\begin{equation} \label{eq:lagrange_dyn}
\textstyle    H(\bm{w}) \bm{\ddot{w}} + \dot{H}(\bm{w})\bm{\dot{w}} - \frac{1}{2} \begin{bmatrix}
    \bm{\dot{w}}^\top \frac{\partial H(\bm{w})}{\partial \bm{w_1}} \bm{\dot{w}} &
    \ldots &
    \bm{\dot{w}}^\top \frac{\partial H(\bm{w})}{\partial \bm{w_n}} \bm{\dot{w}}
\end{bmatrix}^\top = H(\bm{w}) \bm{\ddot{w}} + \bm{c}(\bm{w}, \bm{\dot{w}}) = \bm{\tau}, 
\end{equation}
where $\bm{c}(\bm{w}, \bm{\dot{w}}) \in \reals^{12}$ consists of the coriolis and centrifugal forces. Finally, we apply equation~\eqref{eq:lagrange_dyn} to the servicer satellite states $\bm{w}$ to derive
\begin{equation} \label{eq:First_dynamics}
    \textstyle\begin{bmatrix}
        H_{\textbf{V}} & H_{\textbf{V} {\bm{\Omega}}} & H_{\textbf{V} {\bm{\theta}}} & H_{\textbf{V}\bm{\phi}} \\
        H_{\textbf{V} {\bm{\Omega}}}^\top & H_{{\bm{\Omega}}} & H_{{\bm{\Omega}} {\bm{\theta}}} & H_{{\bm{\Omega}} \bm{\phi}} \\
        H_{\textbf{V} {\bm{\theta}}}^\top & H_{{\bm{\Omega}} {\bm{\theta}}}^\top & H_{{\bm{\theta}}} & H_{{\bm{\theta}} \bm{\phi}} \\
        H_{\textbf{V}\bm{\phi}}^\top & H_{{\bm{\Omega}} \bm{\phi}}^\top & H_{{\bm{\theta}} \bm{\phi}}^\top & H_{\bm{\phi}}
    \end{bmatrix}
    \begin{bmatrix}
        \bm{\dot{v}_B} \\ \bm{\dot{\omega}_B} \\ \bm{\ddot{\theta}} \\ \bm{\ddot{\phi}}
    \end{bmatrix} +
    \begin{bmatrix}
        \bm{c_V} \\ \bm{\bar{c}_b} \\ \bm{\bar{c}_m} \\ \bm{c_r}
    \end{bmatrix} =
    \begin{bmatrix}
        \bm{f_{bV}} \\ \bm{f_{b\Omega}} \\ \bm{\tau_m} \\ \bm{\tau_r}
    \end{bmatrix} \in \reals^{12}.
\end{equation}
Since no external force acts on the system, we set $\bm{f_{bV}} = \bm{f_{b\Omega}} =\bm{0}$. 
Next, we reduce the number of variables involved by solving for $\dot{\bm{v}}_B$ and substituting it in the rest of the equations. In this way we implicitly account for the linear displacement through the coriolis vector and the dependencies of the other equations on the $H_V, \, H_{\textbf{V} {\bm{\Omega}}}, \, H_{\textbf{V} {\bm{\theta}}} \text{ and } H_{\textbf{V}\bm{\phi}}$ matrices. 
We can rewrite the reduced dynamics in terms of generalized inertia matrices $M_b, \, M_{bm}, \, M_{br}, \, M_m, \, M_r$ and nonlinear terms $\bm{c_b}, \, \bm{c_m}, \, \bm{c_r}$. Their explicit expressions in terms of the $H$ matrices are given in Appendix~\ref{APP:inertia_deriv}. 
Using this notation, the dynamics of $[\bm{\dot{\omega}_B}, \, \dot{\bm{\theta}}, \, \dot{\bm{\phi}}]$ can be written compactly as 
\begin{equation} \label{eq:generalized_inertia_M}
    \begin{bmatrix}
        M_b & M_{bm} & M_{br} \\
        M_{bm}^\top & M_m & 0_{3 \times 3} \\
        M_{br}^\top & 0_{3 \times 3} & M_r
    \end{bmatrix}
  \begin{bmatrix}
    \bm{\dot{\omega}_B} \\ \bm{\ddot{\theta}} \\ \bm{\ddot{\phi}}
  \end{bmatrix}
  + \begin{bmatrix}
        \bm{c_b} \\ \bm{c_m} \\ \bm{c_r}
    \end{bmatrix}
  =
  \begin{bmatrix}
    \bm{0}_{3} \\ \bm{\tau_m} \\ \bm{\tau_r}
  \end{bmatrix}.
\end{equation}

To further reduce the complexity of the servicer satellite's dynamics representation, we solve directly for the RW cluster's net acceleration $\bm{\ddot{\phi}} \in \reals^3$ and eliminate it as an additional state. Using the generalized inertia matrices in equation~\eqref{eq:generalized_inertia_M}, we obtain $\bm{\ddot{\phi}} = M_r^{-1} \left( \bm{\tau_r} - M_{br}^\top \bm{\dot{\omega}_B} - \bm{c_r} \right)$.
Therefore, the servicer satellite's dynamics are given by
\begin{equation} \label{eq:system_dynamics}
\begin{bmatrix}
\Tilde{M}_b & \Tilde{M}_{bm} \\
M_{bm}^\top & M_m
\end{bmatrix}
\begin{bmatrix}
\bm{\dot{\omega}_B}(t) \\
\bm{\ddot{\theta}}(t)
\end{bmatrix}
+
\begin{bmatrix}
\bm{\Tilde{c}_b}(t) \\
\bm{c_m}(t)
\end{bmatrix}
=
\begin{bmatrix}
\bm{\tau_r}(t) \\
\bm{\tau_m}(t)
\end{bmatrix} \in \reals^6,
\end{equation}
where $ \tilde{M}_{bm} = - M_{r} M_{br}^{-1} M_{bm}$, $ \bm{\tilde{c}_b} = \bm{c_r} - M_{r} M_{br}^{-1} \bm{c_b} $, and $\tilde{M}_b = M_{br}^T -M_{r} M_{br}^{-1} M_b. $

By the conservation law of angular momentum, any acceleration of the RW cluster redistributes angular momentum between the RW and the satellite base. Equation~\eqref{eq:generalized_inertia_M} shows that the satellite base's acceleration $\bm{\dot{\omega}_B}$ is coupled to the RW cluster acceleration $\bm{\ddot{\phi}}$ and, thus, RW acceleration in one direction generates an equal and opposite rate of change in the satellite base's angular momentum. When the manipulation arm is fixed, an increase in the angular velocity $\bm{\dot{\phi}}$ causes an increase of the angular velocity $\bm{\omega_B}$ with opposite sign, as indicated by the coupling matrix $M_{br}$.
Although not stated explicitly, we compute the remaining coriolis and centrifugal components from the fully expanded version of the dynamics in equation~\eqref{eq:First_dynamics}. Additionally, we note that the generalized inertia matrices $M$ are \emph{state-dependent}, such that the servicer satellite dynamics in equation~\eqref{eq:system_dynamics} remain nonlinear but are control-affine.

\section{MPC framework}
\label{sec:MPC_framework}
In this section, we describe the MPC controller design used for the two phases: A) spin synchronization of the servicer satellite with the target satellite's rotational motion, and B) zero-impulse contact with the target satellite at a designated contact point. We refer to these phases as phase A and B respectively in later sections.

MPC provides a framework that explicitly accounts for the state and actuation constraints of these phases while optimizing the system trajectory over a finite time horizon~\cite{Allgower1998}. 
Specifically, the MPC operates as follows: at each sampling step, it predicts the future system states over a finite horizon and solves a finite horizon open-loop optimal control problem (OCP) subject to the system dynamics and constraints. Rather than executing the entire control sequence generated by the OCP, MPC applies the first control input, measures the resulting system state changes, and repeats this process~\cite{holkar2010overview}. 
This receding-horizon control approach embeds state feedback into the OCP, so that while the optimization runs in open-loop, the overall controller yields a closed-loop solution~\cite{Allgower1998}. MPC handles constraints, optimizes performances, and guarantees infinite horizon stability under appropriately chosen terminal costs and terminal region constraints~\cite{chen1998quasi, findeisen2002introduction}.


\subsection{Spin synchronization MPC}
\label{sec:Phase_A_MPC}
In phase A, we formulate an MPC framework that controls the RW cluster to align both the angular velocity and the orientation between the servicer satellite and the target satellite within the time horizon $\mc{T}_A = [t_{0,A}, t_{0,A} + T_A]$.
During this phase, we assume the manipulation arm's actuation is inactive and the joints are locked at constant angles such that $\bm{\theta}(t) = \bm{\theta}_0 \in \reals^3$, $\bm{\dot{\theta}}(t) = \bm{0}_{3\times1} \in \reals^3$, and $\bm{\ddot{\theta}}(t) = \bm{0}_{3\times1} \in \reals^3$ throughout. We assume the target satellite has a \emph{known} angular velocity $\bm{\omega_{B,\textbf{ref}}} = \bm{\omega_S} \in \reals^3$. Phase A thus aims to drive the servicer satellite's relative orientation quaternion $\bm{q_{\textbf{rel}}}$ to the identity quaternion $\bm{q_f}=[0.0, 0.0, 0.0, 1.0] \in \reals^4$ and its angular velocity $\bm{\omega_B}$ to $\bm{\omega_{B,\textbf{ref}}}$ by the end of this phase. 
To achieve this objective while expending minimal control effort, we use the following objective,
\begin{equation} \label{eq:OCP_1}
\int_{t_{0,A}}^{t_{0,A} + T_A} \bigl(\|\bm{q_{\textbf{rel}}}(t) - \bm{q_f}\|_{Q_{\bm{q}}}^2 + \|\bm{\omega_B}(t) - \bm{\omega_{B,\textbf{ref}}}\|_{Q_{\bm{\omega}}}^2 + \|\bm{\tau_r}(t)\|_{R_r}^2\bigr) dt, \quad \bm{q_f} = [0, 0, 0, 1]. 
\end{equation} 
 where $Q_{\bm{\omega}} \in \reals^{3 \times 3}$ is a  positive definite (PD) matrix for weighing the servicer satellite angular velocity error, 
$Q_{\bm{q}} \in \reals^{4 \times 4}$ is a PD  matrix for weighing the servicer satellite quaternion orientation error, 
and $R_r \in \reals^{3 \times 3}$ is a PD  matrix for weighing the torque applied by the RW cluster. 

When the manipulation arm's actuation is inactive, the servicer satellite's dynamics in equation~\eqref{eq:system_dynamics} take the form $\Tilde{M}_b \bm{\dot{\omega}_B}(t) + \bm{\Tilde{c}_b}(t) = \bm{\tau_r}(t)$. The complete set of constraints that capture the system dynamics, state constraints, and actuation constraints are given by
\begin{align}
    & \bm{\dot{q}}(t) = \frac{1}{2} \begin{bmatrix} 
        -[\bm{\omega_{\mathrm{rel}}} \times] & \bm{\omega_{\mathrm{rel}}} \\
        -\bm{\omega_{\mathrm{rel}}}^\top & 0 
    \end{bmatrix} \bm{q_{\textbf{rel}}}(t),  \ \forall t \in  \mc{T}_A,\label{eqn:quaternion_constraint} \\
    & \bm{\omega_{\mathrm{rel}}}(t) = \bm{\omega_B}(t) - A(\bm{q_{\textbf{rel}}}(t)) \bm{\omega_S},  \ \forall t \in  \mc{T}_A,\label{eqn:rel_ang_vel} \\
    & \bm{\dot{\omega}_B}(t) = \Tilde{M}_b^{-1} \left( \bm{\tau_r}(t) - \bm{\Tilde{c}_b}(t) \right),  \ \forall t \in  \mc{T}_A,\label{eqn:omega_b_ode} \\
    & \bm{\tau_{r,\min}} \leq \|\bm{\tau_r}(t)\|_\infty \leq \bm{\tau_{r,\max}},  \ \forall t \in  \mc{T}_A,\label{eqn:tau_r_bounds} \\
    & \bm{\omega_{B,\min}} \leq \|\bm{\omega_B}(t)\|_2 \leq \bm{\omega_{B,\max}},  \ \forall t \in  \mc{T}_A,\label{eqn:omega_b_bounds} \\
    & \|\bm{q_{\textbf{rel}}}(t_{0,A} + T_A) - \bm{q_f}\|_{Q_{\bm{q}}}^2 \leq \epsilon_{\bm{q}},\label{eqn:quat_term_const} \\
    & \|\bm{\omega_B}(t_{0,A} + T_A) - \bm{\omega_{B,\textbf{ref}}}\|_{Q_{\bm{\omega}}}^2 \leq \epsilon_{\bm{\omega}}.\label{eqn:omega_b_term_const}
\end{align}

The quaternion kinematics in equation~\eqref{eqn:quaternion_constraint} describe the evolution of the servicer satellite's relative orientation based on the relative angular velocity in equation~\eqref{eqn:rel_ang_vel} between the servicer and target satellite. The angular velocity dynamics in equation~\eqref{eqn:omega_b_ode} govern the evolution of the servicer satellite's angular velocity under applied control torques from the moment-generation module. The actuation and state constraints in equations~\eqref{eqn:tau_r_bounds} and~\eqref{eqn:omega_b_bounds} enforce bounds on the RW cluster torques and servicer satellite angular velocity throughout the trajectory to prevent actuator saturation and ensure operational safety, respectively. Finally, the terminal constraints in equations~\eqref{eqn:quat_term_const} and~\eqref{eqn:omega_b_term_const} enforce convergence to the terminal state within a tolerance band of $\epsilon_q \in \reals$ and $\epsilon_\omega \in \reals$, respectively, ensuring acceptable steady-state error without requiring exact convergence.
\subsection{Zero-impulse contact MPC}
\label{sec:Phase_B_MPC}
In phase B, we formulate an MPC that guides the manipulation arm's end effector to the designated contact point on the target satellite while preserving the achieved spin synchronization from phase A. This phase has the time horizon $\mc{T}_B = [t_{0,B}, t_{0,B} + T_B]$, and we assume the manipulation arm's actuation is active and the joints are unlocked. We assume that the designated contact point on the target satellite $\bm{p_{ee}}$ modeled by equation~\eqref{eq:fk_position} is \emph{known} with associated joint angles $\bm{\theta_f} \in \reals^3$. Additionally, we assume that the desired contact velocity $\bm{v_{ee}}$ modeled by equation~\eqref{eq:fk_velocity} is also known with associated joint velocities $\bm{\dot{\theta}_f} \in \reals^3$. Phase B thus aims to smoothly drive the manipulation arm's joint angles $\bm{\theta}$ to $\bm{\theta_f}$ and its joint velocities $\bm{\dot{\theta}}$ to $\bm{\dot{\theta}_f}$ by the end of this phase, while maintaining the servicer satellite’s attitude and angular velocity close to phase A's reference. To reach the desired joint configuration, we use the fifth-order polynomial profile in~\cite{Aghili2024} that smoothly transitions the joints from initial joint values $\bm{\theta_0}$ to final joint values $\bm{\theta_{f}}$ given by  
\begin{align}
    \begin{bmatrix}
        \bm{\theta}_{\textbf{ref}}(t) \\
        \bm{\dot{\theta}}_{\textbf{ref}}(t) \\
        \bm{\ddot{\theta}}_{\textbf{ref}}(t)
    \end{bmatrix} &=
    \begin{bmatrix}
        \bm{\theta_0} + (3\hat{t}^2 - 2\hat{t}^3)\Delta\bm{\theta} \\
        (6\hat{t} - 6\hat{t}^2)\Delta\bm{\theta}/t_f \\
        (6 - 12\hat{t})\Delta\bm{\theta}/t_f^2
    \end{bmatrix}, \,
    \hat{t} = \frac{t}{t_f}, \, \Delta\bm{\theta} = \bm{\theta_f} - \bm{\theta_{0}}, t_f
    = \max \left\{
        \frac{3}{2\norm{\bm{\dot{\theta}_{\max}}}}_2 \max_j |\Delta\theta_j|,
        \left( \frac{6}{\norm{\bm{\ddot{\theta}_{\max}}}}_2 \max_j |\Delta\theta_j| \right)^{1/2}
    \right\},
\label{eq:spline_formula}
\end{align} where $\bm{\theta}_{\textbf{ref}}(t)$, $\bm{\dot{\theta}}_{\textbf{ref}}(t)$, and $\bm{\ddot{\theta}}_{\textbf{ref}}(t) \in \reals^3$ are the time-varying spline trajectories for the joint angles, velocities, and accelerations. The joint angle spline trajectory $\bm{\theta}_{\textbf{ref}}(t)$ and the joint velocity $\bm{\dot{\theta}}_{\textbf{ref}}(t)$ are used in the objective function for phase B to ensure the joints smoothly move from the initial configuration $\bm{\theta_0}$ to the desired final joint configuration $\bm{\theta_f}$. The term $t_f$ denotes the final time where the manipulation arm reaches the desired joint angle configuration and $\Delta \theta_j$ is the displacement in joint angles for $j \in [1,3]$.


Phase B's MPC has the following objective,
\begin{equation} \label{eq:OCP_2}
    \int_{t_{0,B}}^{t_{0,B} + T_B}  \|\bm{\tau_r}(t)\|^2_{R_r} + \|\bm{\tau_m}(t)\|^2_{R_m} + \|\bm{\theta}(t) - \bm{\theta}_{\textbf{ref}}(t)\|^2_{Q_{\bm{\theta}}} + \|\bm{\dot{\theta}}(t) - \bm{\dot{\theta}}_{\textbf{ref}}(t)\|^2_{\widehat{Q}_{{\bm{\theta}}}} + \|\bm{q_{\textbf{rel}}}(t) - \bm{q_f}\|^2_{Q_{\bm{q}}} + \|\bm{\omega_B}(t) - \bm{\omega_{B,\textbf{ref}}}\|^2_{Q_{\bm{\omega}}} \, dt,
\end{equation} 
where $Q_{\bm{\omega}}, \, Q_{\bm{q}}, \, R_r$ are similarly defined as in equation~\eqref{eq:OCP_1}. Additionally,  we introduce three new weight matrices corresponding to the manipulation module: $Q_{\bm{\theta}} \in \reals^{3 \times 3}$, a PD matrix for weighing the manipulation module joint angle error, $\widehat{Q}_{{\bm{\theta}}} \in \reals^{3 \times 3}$, a PD matrix for weighing the manipulation module joint velocity error, and $R_m \in \reals^{3 \times 3}$, a PD matrix for weighing the torque generated by the manipulation module arm.


In addition to the phase A's constraints listed in equations~\eqref{eqn:quaternion_constraint},~\eqref{eqn:rel_ang_vel},~\eqref{eqn:tau_r_bounds},~\eqref{eqn:omega_b_bounds},~\eqref{eqn:quat_term_const}, and~\eqref{eqn:omega_b_term_const}, phase B's MPC constraints are
\begin{align}
    & \begin{bmatrix}
    \bm{\dot{\omega}_B}(t) \\
    \bm{\ddot{\theta}}(t)
    \end{bmatrix}
    =
    \begin{bmatrix}
    \Tilde{M}_b & \Tilde{M}_{bm} \\
    M_{bm}^\top & M_m
    \end{bmatrix}^{-1} \left(\begin{bmatrix}
    \bm{\tau_r}(t) \\
    \bm{\tau_m}(t)
    \end{bmatrix} - \begin{bmatrix}
    \bm{\Tilde{c}_b}(t) \\
    \bm{c_m}(t)
    \end{bmatrix} \right),  \ \forall t \in  \mc{T}_B\label{eq:coupled_ode}\\
    & \bm{\tau_{m,\min}} \leq \norm{\bm{\tau_m}(t)}_\infty \leq \bm{\tau_{m,\max}},  \ \forall t \in  \mc{T}_B, \label{eq:tau_m_bounds}\\
    & \bm{\theta_{\min}} \leq \norm{\bm{\theta}(t)}_2 \leq \bm{\theta_{\max}},  \ \forall t \in  \mc{T}_B,\label{eq:theta_bounds}\\
    & \bm{\dot{\theta}_{\min}} \leq \norm{\bm{\dot{\theta}}(t)}_2 \leq \bm{\dot{\theta}_{\max}},  \ \forall t \in  \mc{T}_B,\label{eq:theta_dot_bounds}\\
    & \|\bm{\theta}(t_{0,B} + T_B) - \bm{\theta}_{\textbf{ref}}\|_{Q_{\bm{\theta}}}^2 \leq \epsilon_{\bm{\theta}},  \label{eq:theta_term_const}\\
    & \|\bm{\dot{\theta}}(t_{0,B} + T_B) - \bm{\dot{\theta}}_{\textbf{ref}}\|_{\widehat{Q}_{{\bm{\theta}}}}^2 \leq \hat{\epsilon}_{{\bm{\theta}}}. \label{eq:theta_dot_term_const}
\end{align} The momentum-coupled dynamics in equation~\eqref{eq:coupled_ode} govern the evolution of both the servicer satellite's angular velocity and the manipulator joint accelerations under the combined influence of the RW cluster torques and manipulator joint torques, as derived in Section~\ref{sec:moment_couple_dynamics}. State and actuation constraints in equations~\eqref{eq:tau_m_bounds},~\eqref{eq:theta_bounds}, and~\eqref{eq:theta_dot_bounds} enforce bounds on the manipulator joint torques, joint angles, and joint velocities throughout the trajectory to prevent the joint actuators from exceeding their power limits and ensure kinematic feasibility. Terminal constraints in equations~\eqref{eq:theta_term_const} and~\eqref{eq:theta_dot_term_const} enforce convergence to the terminal state within a tolerance band of $\epsilon_{{\bm{\theta}}} \in \reals$ and $\hat{\epsilon}_{{\bm{\theta}}} \in \reals$, respectively, ensuring acceptable steady-state error without requiring exact convergence.

\section{Simulation and results}\label{sec:sim}
In this section, we implement the nonlinear MPC solver formulated in Section~\ref{sec:MPC_framework} and compare its performance  to PID controller approaches suggested in prior work~\cite{Aghili2019}.  
To solve equations~\eqref{eq:OCP_1} and~\eqref{eq:OCP_2}, we use the nonlinear MPC solver \texttt{acados}~\cite{acados}, a software package that provides fast and embedded solvers  to conduct the simulations. \texttt{acados} discretizes the continuous-time dynamics as described in equation~\eqref{eq:system_dynamics} by using Runge-Kutta (RK) integration method, and the continuous-time OCP in equations~\eqref{eq:OCP_1} and~\eqref{eq:OCP_2} by using multiple shooting methods~\cite{acados}.
To implement the PID controller, we describe its structure and tuning in Section~\ref{sec:PID_framework}.
The specific MPC configurations we use in  our results are provided in Appendix~\ref{APP:_MPC_solver} and~\ref{APP:solve_params}.
Over both operation phases, we compare PID and MPC controller's performance in terms of constraint violation, tracking error, maneuver time, and operation success rate. 
\subsection{Environment setup}
For all case studies, we simulate the following two phases: spin synchronization (phase A) and zero-impulse contact (phase B), as described in Sections~\ref{sec:Phase_A_MPC} and~\ref{sec:Phase_B_MPC}. Each phase executes for a maximum duration of $75$ seconds, with a sampling time of $\Delta t = 0.01$ second and a prediction horizon length of $T_A = T_B = 0.7$ seconds. Each case study consists of $50$ randomized MC trials conducted under the conditions listed in Section~\ref{sec:case_study_setup}. We then plot out the results of our analysis to illustrate the comparison between the MPC and PID controllers.

\textbf{Actuation saturation}. We assume that the actuation modules are power limited, therefore, all torque values requested by the controller, $\bm{\tau_{r,\textbf{cmd}}}$, are saturated element-wise prior to execution, such that the executed torques, $\bm{\tau_{r,\textbf{actual}}}, \ \bm{\tau_{m,\textbf{actual}}}$, satisfy
\begin{equation}
\bm{\tau_{r,\textbf{actual}}} = \text{sat}(\bm{\tau_{r,\textbf{cmd}}}, \bm{\tau_{r,\min}}, \bm{\tau_{r,\max}}),\quad \bm{\tau_{m,\textbf{actual}}} = \text{sat}(\bm{\tau_{m,\textbf{cmd}}}, \bm{\tau_{m,\min}}, \bm{\tau_{m,\max}}),
\end{equation} 
where the saturation function is element-wise defined as $\bigl[\text{sat}(\bm{\tau}, \bm{\tau_{\min}}, \bm{\tau_{\max}})\bigr]_i = \min\bigl\{\bm{\tau_{\max, i}}, \max\{\bm{\tau_i}, \bm{\tau_{i,\min}}\}\bigr\} \forall i \in [3]$. 
This ensures $\|\bm{\tau_{r,\textbf{actual}}}\|_\infty \leq \bm{\tau_{r,\max}}$ and $\|\bm{\tau_{m,\textbf{actual}}}\|_\infty \leq \bm{\tau_{m,\max}}$ and that the commanded torques remain within physically realizable bounds for both the MPC and PID controllers. We denote the values of the minimum and maximum torques as equivalent to the following
\begin{align}
    \bm{\tau_{r,\min}} = [-2, -2, -2] \text{ (N $\cdot$ m)}, &\quad \bm{\tau_{r,\max}} = [2, 2, 2]\text{ (N $\cdot$ m)},\label{eq:max_torques_r}\\
    \bm{\tau_{m,\min}} = [-0.3,-0.3,-0.3]\text{ (N $\cdot$ m)}, &\quad \bm{\tau_{m,\max}}=[0.3,0.3,0.3]\text{ (N $\cdot$ m)}.\label{eq:max_torques_m}
\end{align}

\textbf{Dynamics uncertainty}. The satellite dynamics are modeled by equation~\eqref{eq:system_dynamics}
with numerical values provided in Appendix~\ref{APP:sim_params}. We assume that the operating system parameters differ from the nominal system parameters used in the MPC/PID controllers in our MC trials. Specifically, each scalar physical parameter $p$ in the model is perturbed by $10\%$ Gaussian uncertainty with standard deviation $\sigma_p \in \reals$. This implies that each operating system parameter $p_{\text{true}} \in \reals$ is derived from the nominal parameter values $p_{\text{nom}} \in \reals$ as
\begin{equation} \label{eq:gaussian_uncertainty}
    p_{\text{true}} \sim \mathcal{N}\left(p_{\text{nom}}, \text{diag}(\sigma_p)\right), \quad \sigma_p = 0.1 \cdot p_{\text{nom}}. 
\end{equation}
This rule is applied element-wise to all masses $m_i$ and link lengths $L_1, \, L_2, \, L_3$, as well as to each entry of the moment-generation base inertia matrix $I_b = \begin{bmatrix}
    I_{b,1} & I_{b,2} & I_{b,3} \\
    I_{b,4} & I_{b,5} & I_{b,6} \\
    I_{b,7} & I_{b,8} & I_{b,9}
\end{bmatrix} \in \reals^{3\times3}$. The individual components of the moment-generation base matrix $I_{b,i} \ \forall \ i \in [9]$ are resampled according to equation~\eqref{eq:gaussian_uncertainty} until the reconstructed matrix $I_b$ has strictly positive eigenvalues, i.e., all $\lambda_b>0, \, \lambda_b \in \reals, \, \forall \ b \in [1,3]$, meaning that the moment-generation base matrix is PD. The remaining inertia matrices (for the manipulation arm and the RW cluster) are not sampled element-wise, but they are recomputed from the sampled masses, link lengths, and geometric parameters using the formulas in Appendix~\ref{APP:sim_params}.
Initial system states as well as the nominal terminal state values $\bm{\omega_{B,\textbf{ref}}}, \, \bm{q_f}, \, \bm{\theta_f}, \bm{\dot{\theta}_f}$ are perturbed using Gaussian distributions with state‑dependent standard deviations that are $10\%$ of their nominal parameter values. For example, for the initial servicer satellite angular velocities, we use
$\bm{\omega}_{\bm{B,0},\textbf{true}} \sim \mathcal{N}\big(\bm{\omega}_{\bm{B,0},\textbf{nom}}, \text{diag}(\bm{\sigma_\omega})\big), $ where $\bm{\sigma_\omega} \in \reals^3$ is the state-dependent standard deviation computed element-wise as $\sigma_{\omega,i} = 0.1 \cdot |\omega_{B,0,i}| + \sigma_{f,\omega} \ \forall \ i \in  [1,3]$, and $\sigma_{f,\omega} = 10^{-2}$ is an additive floor value that prevents stability issues when nominal state values are zero.

\textbf{Simulation vs control synthesis}. The controllers use the true parameters to calculate the dynamics as specified in equation~\eqref{eq:system_dynamics}, and the true parameters are utilized to solve the optimal control problem described in equations~\eqref{eq:OCP_1} and~\eqref{eq:OCP_2} respectively for our MPC framework, while the control for our PID controller is generated by equations later described in Section~\ref{sec:PID_framework}.
The nominal parameters for the servicer satellite are specified in Table~\ref{tab:parameters}, and the nominal states for each phase are specified in Section~\ref{sec:ref_traj_def}. 

\subsection{Reference trajectory definition}
\label{sec:ref_traj_def}
We define the reference state values used in the MPC controller and PID controller. 

\textbf{Phase A: spin synchronization.}
Phase A aims to generate RW torques $\bm{\tau_r}$ modeled by equation~\eqref{eq:rw_torque} to drive the servicer satellite's angular velocity $\bm{\omega_B}$ to a desired angular velocity $\bm{\omega_{B,\textbf{ref}}} = \bm{\omega_S}$ modeled by equation~\eqref{eq:omega_rel}, the relative orientation quaternion $\bm{q_{\textbf{rel}}}$ to the desired quaternion $\bm{q_{f}}$, while minimizing the total RW torque expended. For all case studies, we use the reference values
\begin{equation}\label{eq:spin_sync_refs}
   \bm{\omega_{B,\textbf{ref}}}= [0.0, 0.0, 0.2] \text{ (rad / s)}. 
\end{equation}
These values also define phase A's MPC objective in equation~\eqref{eq:OCP_1}. Furthermore, throughout phase A, we assume that the manipulation arm joint angles are locked at 
\begin{equation}
\bm{\theta_{0}} = [0.05, 0.4, 0.05] \text{ (rad)}.
\end{equation}

\textbf{Phase B: zero-impulse contact.}
Phase B aims to generate both RW torques $\bm{\tau_r}$ modeled by equation~\eqref{eq:rw_torque} and manipulator joint torques $\bm{\tau_m}$ by equation~\eqref{eqn:joint_torques} so that the servicer satellite can make zero-impulse contact with the target satellite at the designated contact points. During this phase, we maintain phase A's references for the servicer satellite moment-generation base~\eqref{eq:spin_sync_refs}. Additionally, the manipulator joint angles have desired terminal values given by 
\begin{equation}\label{eq:terminal_joint_angles}
     \bm{\theta_{f}} = [0.5, 0.2, 0.3] \text{ (rad)}, \quad \bm{\dot{\theta}_{f}} = [0.0, 0.0, 0.0] \text{ (rad)}. 
\end{equation}
To reach the desired joint configurations, we use the reference joint trajectories $\bm{\theta}_{\textbf{ref}}(t)$ and $\bm{\dot{\theta}}_{\textbf{ref}}(t)$ as described in equation~\eqref{eq:spline_formula} to prevent sudden aggressive jerking motion in the manipulation module. We note that phase B's initial joint angle values $\bm{\theta_0} \in [0, 2\pi]^3$ depend on phase $A$. The desired final joint configuration $\bm{\theta_f}$ varies based on case study, and we provide specific numerical values within each case study.


\subsection{PID controller design and tuning}
\label{sec:PID_framework}

To evaluate the performance improvements of the MPC controller, we establish a PID baseline expanded off a control law proposed by~\cite{Aghili2024}. We introduce the control term $\bm{u}_{\textbf{att}}(t)$ defined as \begin{equation}
    \bm{u}_{\textbf{att}}(t) = k_{\bm{q}} \bm{q_{\upsilon,\textbf{rel}}}(t) + k_{\bm{\omega}} \bm{\omega_{\textbf{rel}}}(t) + k_{i,{\bm{q}}} \int_{t_{0,A}}^{t_{0,A}+T_A} \bm{q_{\upsilon,\textbf{rel}}}(t) dt + k_{i,{\bm{\omega}}} \int_{t_{0,A}}^{t_{0,A}+T_A} \bm{\omega_{\textbf{rel}}}(t) dt  + k_{d,{\bm{q}}}\bm{\dot{q}_{\upsilon,\textbf{rel}}}(t) + k_{d,{\bm{\omega}}} \bm{\dot{\omega}_\textbf{rel}}(t),
\end{equation}
where the feedback gains $k_{\bm{q}}, \, k_{\bm{\omega}}, \, k_{i,{\bm{q}}}, \, k_{i,{\bm{\omega}}}, \, k_{d,{\bm{q}}}, \, k_{d,{\bm{\omega}}}  \in \reals_+$ are the proportional, integral, and derivative gains for attitude and angular velocity control, respectively.
We then utilize $\bm{u}_{\textbf{att}}(t) \in \reals^3$ in the PID feedback law for phase A as follows
\begin{equation} \label{eq:PID_1}
    \bm{\tau_r}(t) = \bm{\tilde{c}_b}(t) - \Tilde{M}_b\bm{u}_{\textbf{att}}(t), \ \forall t \in \mc{T}_A.
\end{equation} The controller minimizes the relative orientation quaternion vector $\bm{q_{\upsilon,\textbf{rel}}}$ in equation~\eqref{eq:rel_quat} and relative angular velocity $\bm{\omega_{\textbf{rel}}}$ in equation~\eqref{eq:omega_rel} such that spin synchronization is achieved when
\begin{equation}
    \bm{q_{\upsilon,\textbf{rel}}}(t) = \bm{0}_{3\times1}, \quad \bm{\omega_{\textbf{rel}}}(t) = \bm{0}_{3\times1}.
\end{equation}

In Phase B, the manipulation arm joints are unlocked and momentum coupling between the moment-generation module and manipulation module appears explicitly. We introduce the control term $\bm{u}_{\textbf{arm}}(t) \in \reals^{3}$, and redefine the previously used control term $\bm{u}_{\textbf{att}}(t)$ for phase B as follows
\begin{equation}
\bm{u}_{\textbf{att}}(t) = k_{\bm{q}} \bm{q_{\upsilon,\textbf{rel}}}(t) + k_{\bm{\omega}} \bm{\omega_{\textbf{rel}}}(t) + k_{i,{\bm{q}}} \int_{t_{0,B}}^{t_{0,B}+T_B} \bm{q_{\upsilon,\textbf{rel}}}(t) dt + k_{i,{\bm{\omega}}} \int_{t_{0,B}}^{t_{0,B}+T_B} \bm{\omega_{\textbf{rel}}}(t) dt  + k_{d,{\bm{q}}}\bm{\dot{q}_{\upsilon,\textbf{rel}}}(t) + k_{d,{\bm{\omega}}} \bm{\dot{\omega}_\textbf{rel}}(t),
\end{equation}
\begin{equation}
\bm{u}_{\textbf{arm}}(t) = k_p \big(\bm{\theta}_{\textbf{ref}}(t) - \bm{\theta}(t)\big) + k_i \int_{t_{0,B}}^{t_{0,B}+T_B} \bigl(\bm{\theta}_{\textbf{ref}}(t) - \bm{\theta}(t)\bigr) dt + k_d \bigl(\bm{\dot{\theta}}_{\textbf{ref}}(t) - \bm{\dot{\theta}}(t)\bigr) + \bm{\ddot{\theta}}_{\textbf{ref}}(t),
\end{equation}
where $\bm{\theta}_{\textbf{ref}}(t), \, \bm{\dot{\theta}}_{\textbf{ref}}(t), \, \bm{\ddot{\theta}_{\textbf{ref}}}(t) \in \reals^3$ are the reference trajectories for the joint angles, velocities, and accelerations from equation~\eqref{eq:spline_formula}, and $k_p, \, k_i, \, k_d \in \reals_+$ are the PID gains for joint angle and velocity tracking. 
We combine these control terms with the system dynamics in equation~\eqref{eq:system_dynamics} to derive the PID feedforward law given by
\begin{equation} \label{eq:PID_2}
\begin{bmatrix}
\bm{\tau_r}(t) \\
\bm{\tau_m}(t)
\end{bmatrix}
=
\begin{bmatrix}
\bm{\tilde{c}_b}(t) \\
\bm{c_m}(t)
\end{bmatrix}
-
\begin{bmatrix}
\Tilde{M}_b & \Tilde{M}_{bm} \\
M_{bm}^\top & M_m
\end{bmatrix}
\begin{bmatrix}
\bm{u_{\textbf{att}}}(t) \\
\bm{u_{\textbf{arm}}}(t)
\end{bmatrix}, \quad \forall t \in \mc{T}_B.
\end{equation}
All feedback gains are tuned using the closed-loop Ziegler-Nichols method~\cite{copeland2008design}. Integral and derivative gains are initially set to zero, and the proportional gain is increased until sustained oscillations occur. 
The resulting gain formulas and the numerical values used in our simulations are provided in Appendix~\ref{APP:PID_tuning_details}.


\subsection{Case study set up and performance metrics}
\label{sec:case_study_setup}
This section discusses the case study set up to compare the performance of the proposed MPC controller from Section~\ref{sec:MPC_framework}, with a baseline PID controller from Section~\ref{sec:PID_framework}.

\textbf{Case study A}. This case study compares the controller performance for the nominal setting, where the terminal joint angles and velocities  $\bm{\theta_{f}},\bm{\dot{\theta}_{f}}$~\eqref{eq:terminal_joint_angles} are time-invariant. The terminal joint angle correspond to an  end effector contact point location of $x_\text{ee} = 1.06, \ y_\text{ee} = 1.03, \ \theta_\text{ee} = 1.0 \, \text{rad} \ (57.3^\circ)$. During the control horizon $\mc{T}_B$, the  joint reference trajectories $\bm{\theta}_{\textbf{ref}}(t), \bm{\dot{\theta}}_{\textbf{ref}}(t)$ are derived using the given terminal joint values based on the spline defined in~\eqref{eq:spline_formula}.

\textbf{Case study B}. This case study compares controller performances when the terminal joint angles $\bm{\theta_{f}}, \bm{\dot{\theta}_{f}}$ differ from~\eqref{eq:terminal_joint_angles}, and takes on time-varying values instead. The satellite base state references remain time-invariant, while the joint angles and velocities follow a time-varying reference trajectory that reflects updated contact points that may result from real-time sensor updates. The joint angles must converge to the reference trajectory $\bm{\theta}_{\textbf{ref}}(t) $ and the joint velocities must simultaneously be driven to $\bm{\dot{\theta}_{\textbf{ref}}}(t) $, where
\begin{equation}
    \bm{\theta}_{\textbf{ref}}(t) = \begin{bmatrix}
        A_{\text{nom}} \cos{(B_{\text{nom}}t)} & A_{\text{nom}} \sin{(B_{\text{nom}}t)} & K_{\text{nom}}t
    \end{bmatrix}\in \reals^3,\ \forall t \in \mc{T}_B,
    \label{eq:theta_ref}
\end{equation}
\begin{equation}
    \bm{\dot{\theta}}_{\textbf{ref}}(t)= \begin{bmatrix}
        -A_{\text{nom}}B_{\text{nom}} \sin{(B_{\text{nom}}t)} & A_{\text{nom}}B_{\text{nom}} \cos{(B_{\text{nom}}t)} & K_{\text{nom}}
    \end{bmatrix} \in \reals^3,\ \forall t \in \mc{T}_B.
    \label{eq:theta_dot_ref}
\end{equation}
We use nominal parameter values $A_{nom} = 0.1$, $B_{nom} = 0.5$, and $K_{nom}=0.01$. 

\textbf{Case study C}. This case study compares the controllers' performance under observation and actuation noise. The terminal joint angle values $\bm{\theta_{f}}$ and terminal joint velocities, $\bm{\dot{\theta}_{f}}$, modeled by equation~\eqref{eq:terminal_joint_angles}, are time-invariant and the joint reference trajectories $\bm{\theta}_{\textbf{ref}}(t), \, \bm{\dot{\theta}}_{\textbf{ref}}(t)$ are the same as in case study A. The observation noise is modeled as state-dependent Gaussian noise $\mc{N}\big(0, \text{diag}(\bm{\sigma_{O}})\big)$, with each standard deviation value for each state being defined element-wise as $0.5\%$ of the nominal terminal state values $\bm{\omega_{B,\textbf{ref}}}, \, \bm{q_f}, \, \bm{\theta_f}, \, \bm{\dot{\theta}_f}$ as follows
\begin{equation}
    \sigma_{O,i} = 0.005 \ \cdot |x_i| + \sigma_{f,i}, \ \forall \ i \in [13],
\end{equation} where $\sigma_{f,i} \in \reals$ is an additive floor value that prevents stability issues when the nominal terminal state values are zero. Similarly, the actuation noise is modeled as control-dependent Gaussian noise  $\mc{N}\big(0, \text{diag}(\bm{\sigma_{A}})\big)$, which the standard deviation is element-wise defined  as $0.5\%$ of the maximum torque values modeled by equations~\eqref{eq:max_torques_r} and~\eqref{eq:max_torques_m} as follows \begin{equation}
    \sigma_{A,j} = 0.005 \ \cdot |\tau_j|, \ \forall \ j \in [6].
\end{equation} In phase A, we define the observation noise as $\bm{w}_A \sim \mathcal{N}\bigl(\bm{0}, \text{diag}(\bm{\sigma^A_{O}})\bigr) \in \reals^{7}$, where the standard deviation is $\bm{\sigma^A_{O}} =  10^{-4}\begin{bmatrix}
    1& 1& 10& 1& 1& 1& 50
\end{bmatrix} \in \reals^{7}$, and the actuation noise in phase A is defined as $\bm{n}_A \sim \mathcal{N}\bigl(\bm{0}, \text{diag}(\bm{\sigma^A_{A}})\bigr) \in \reals^3$, where $ \bm{\sigma^A_{A}} = 10^{-2}\begin{bmatrix}
       1 & 1 & 1
    \end{bmatrix}\in \reals^3$.
In phase B, the observation noise is defined as $\bm{w}_B \sim \mathcal{N}\bigl(\bm{0}, \text{diag}(\bm{\sigma^B_O})\bigr) \in \reals^{13}$ and the actuation noise is defined as  $\bm{n}_B \sim \mathcal{N}\bigl(\bm{0}_{6\times1}, \text{diag}(\bm{\sigma^B_A})\bigr) \in \reals^6$. The standard deviations for the observation and actuation noise for phase B are
\begin{align}
    \bm{\sigma^B_O} = 10^{-4}
\Bigl[\,25 \quad 10 \quad 15 \quad 1 \quad 1 \quad 10 \quad 1 \quad 1 \quad 1 \quad 1 \quad 1 \quad 1 \quad 50\,\Bigr]
 \in \reals^{13}, \\
\bm{\sigma^B_A} = 10^{-4}\begin{bmatrix}
        100 & 100 & 100 & 15 & 15 & 15
    \end{bmatrix}\in \reals^6,
\end{align} respectively.
These noises are incorporated into the system dynamics as
\begin{equation}
    \dot{\bm{x}} = f(\bm{x} + \bm{w}, \bm{u} + \bm{n}),
\end{equation} 
where $\bm{n}$ and $\bm{w}$ switch between their phase A and phase B definitions according to the current mission phase, and $f$ are the dynamics which alternate between phase A dynamics modeled by equation~\eqref{eqn:omega_b_ode}, and phase B dynamics modeled by equation~\eqref{eq:coupled_ode}, according to the current mission phase. 

\textbf{Performance metrics}. We use the following metrics to compare controller performance. To simplify notation, we collectively refer to the servicer satellite's states across phase A and phase B as 
$\bm{x} = \begin{bmatrix}
        \bm{\theta} & \bm{\omega_B} & \bm{\dot{\theta}} & \bm{q_{\textbf{rel}}}
    \end{bmatrix} \in \reals^{13}$.

\textbf{Constraint Violation (CV)}.  Violation of state bounds $\bm{x}_{\min}, \, \bm{x}_{\max} \in \reals^{13}$ for a given state $\bm{x}$:
    \begin{equation} \label{eq:Constraint_violation}
        CV(\bm{x}) = 
        \bigl\lVert [\bm{x}_{\min} - \bm{x}]_+ \bigr\rVert_2
        +
        \bigl\lVert [\bm{x} - \bm{x}_{\max}]_+ \bigr\rVert_2,
    \end{equation} where $[\cdot]_+ = \max(0, \cdot)$ takes the element-wise non-negative value of the input vector.
    
\noindent\textbf{Root Mean Square Error $(\text{RMSE})$}. Time-averaged tracking error between the system state $\bm{x}(t)$ and the reference $\bm{x}_{\textbf{ref}}(t) \in \reals^{13}$, given by
    \begin{equation} \label{eq:RMSE_critetion}
        \textstyle\text{RMSE}= \sqrt{\frac{1}{T} \int_{t_0}^T \|\bm{x}(t) - \bm{x}_{\textbf{ref}}(t)\|_2^2 \ dt} .
    \end{equation}
\noindent\textbf{Average Computation Time $\bar{t}_{\text{comp}}$}. Average time required to compute the control input. Let $t_{\text{comp}}(t)$ denote the computation time at time $t$, then the average computation time is given by
\begin{equation}
   \textstyle \bar{t}_{\text{comp}} = \frac{1}{T} \int_{t_0}^T t_{\text{comp}}(t) \ dt.
\end{equation}
\textbf{Convergence criteria and constraints.} In addition to the criteria listed above, we also track state divergence during simulations. We define state divergence to occur at time $t$ when \begin{equation} \label{eq:divergence}
    \textstyle\|\bm{\omega_B}(t) -\bm{\omega_{B,\textbf{ref}}}||_2 \geq \psi, \ \quad \textstyle\|\bm{q}_{\textbf{rel}}(t) -\bm{q_f}||_2 \geq \psi, \ \quad \textstyle\|\bm{\theta}(t) -\bm{\theta_f}||_2 \geq \psi, \ \quad \textstyle\|\bm{\dot{\theta}}(t) -\bm{\dot{{\theta}}_f}||_2 \geq \psi,
\end{equation} where $\psi = 10^{6} \in \reals$ is the divergence criterion for which a state value exceeds physically realizable bounds. $\psi$ is set as an arbitrary large number, and ensures that if the two-norm state differences listed in equation~\eqref{eq:divergence} reach or exceed $\psi$, then this would physically represent catastrophic loss of control of the servicer satellite at time $t$. Furthermore, we define state convergence to occur at time $t$ when \begin{equation} \label{eq:convergence}
   \textstyle\|\bm{\omega_B}(t) -\bm{\omega_{B,\textbf{ref}}}||_2 \leq \xi, \ \quad \textstyle\|\bm{q}_{\textbf{rel}}(t) -\bm{q_f}||_2 \leq \xi, \ \quad \textstyle\|\bm{\theta}(t) -\bm{\theta_f}||_2 \leq \xi, \ \quad \textstyle\|\bm{\dot{\theta}}(t) -\bm{\dot{{\theta}}_f}||_2 \leq \xi,
\end{equation} where $\xi = 10^{-3} \in \reals$ is the convergence criterion for which a state value approaches its reference value. $\xi$ is set as an arbitrary small number, and ensures that if the two-norm state differences listed in equation~\eqref{eq:convergence} reach or exceed $\xi$, then this would physically represent the current mission phase being completed at time $t$.

\subsection{Results and discussion}
\begin{table}[ht!]
\centering
\caption{Case Study Performance Comparison Table}
\label{tab:all_case_studies}
\setlength{\tabcolsep}{3pt}
\renewcommand{\arraystretch}{1.15}
\footnotesize
\begin{tabular}{llccccccc}
\hline
\textbf{Case Study} 
& \textbf{Controller}
& \textbf{$\bm{q_{\textbf{rel},\text{RMSE}}}$}
& \textbf{$\bm{\omega_{B,\text{RMSE}}}$}
& \textbf{$\bm{p_{ee,\text{RMSE}}}$}
& \textbf{$\bm{v_{ee,\text{RMSE}}}$}
& \textbf{$\bar{t}_{\text{comp}}$}
& \textbf{$CV \ \%$}
& \textbf{Success $\%$} \\
& & [rad] & [rad/s] & [m] & [m/s] & [s] & [\%] & [\%] \\
\hline
{\textbf{Case Study A}}
& MPC & {$0.0087$} & $0.00084$ & $0.28$ & $0.0096$ & $1.0$ & $0$ & $86$ \\
& PID & $0.031$ & $0.0073$ & $0.33$ & $0.012$ & $0.018$ & $0.40$ & $64$ \\
\cline{1-9}
{\textbf{Case Study B}}
& MPC & $0.016$ & $0.0033$ & $0.38$ & $0.07$ & $0.72$ & $0$ & $92$ \\
& PID & $0.077$ & $0.018$ & $0.41$ & $0.14$ & $0.019$ & $8.03$ & $80$ \\
\cline{1-9}
{\textbf{Case Study C}}
& MPC &$0.81$ & $0.054$ & $0.30$ & $0.02$ & $0.49$ & $0$ & $60$ \\
& PID & $0.269$ & $0.058$ & $0.95$ & $0.41$ & $0.018$ & $26.41$ & $10$ \\
\hline
\end{tabular}
\vspace{-10pt}
\end{table}
All plots use shaded error bands to denote the interquartile range (IQR) of the respective quantity over random MC trials, and use solid lines to represent the median over the same MC trials. IQR and median are computed in the linear scale and visualized on the logarithmic scale.

\begin{figure}[ht!]
    \centering
    \begin{subfigure}[t]{0.48\linewidth}
        \centering
        \includegraphics[width=\linewidth]{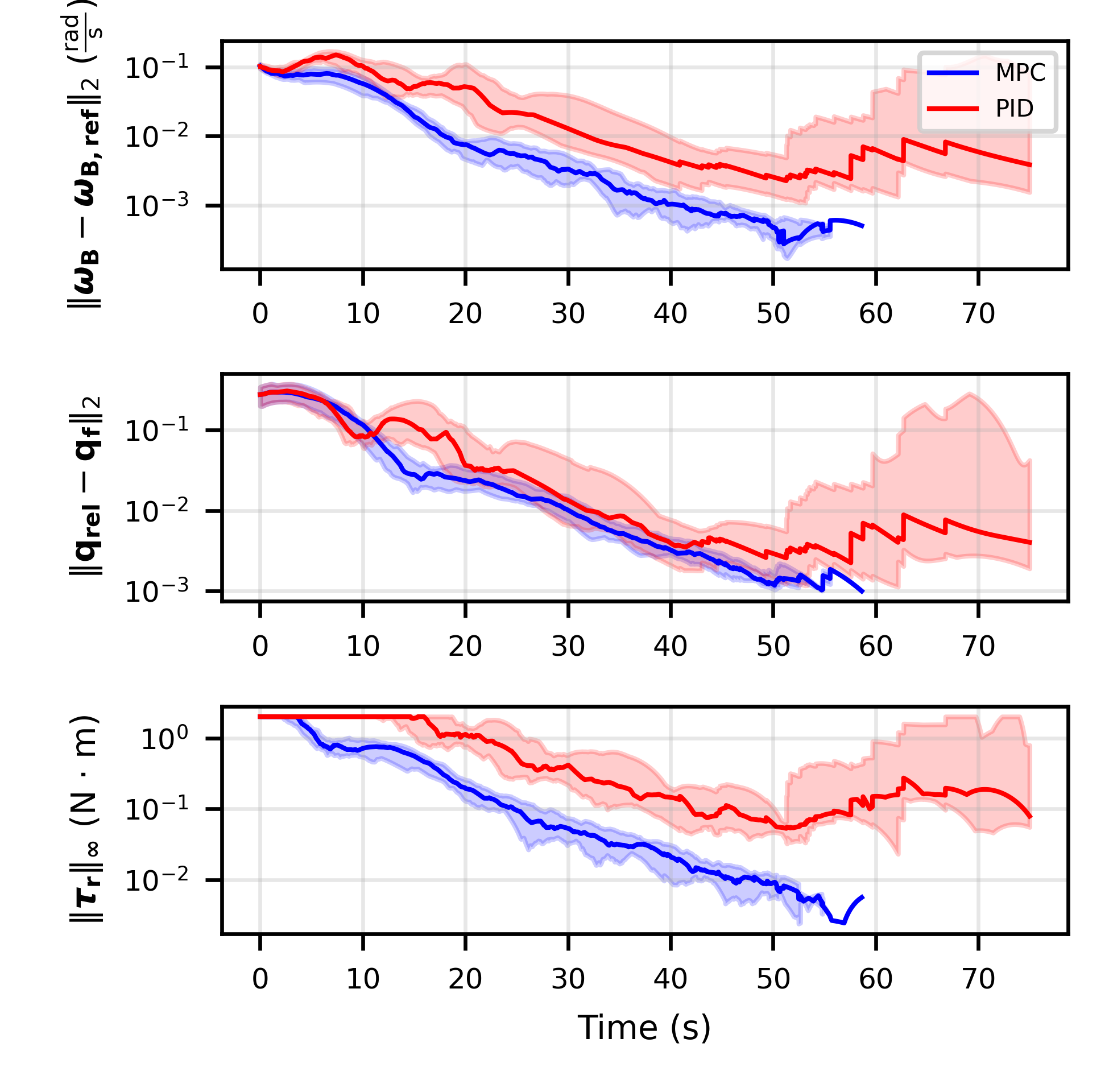}
        \label{fig:mc_case_a_phase_a}
    \end{subfigure}
    \hfill
    \begin{subfigure}[t]{0.48\linewidth}
        \centering
        \includegraphics[width=\linewidth]{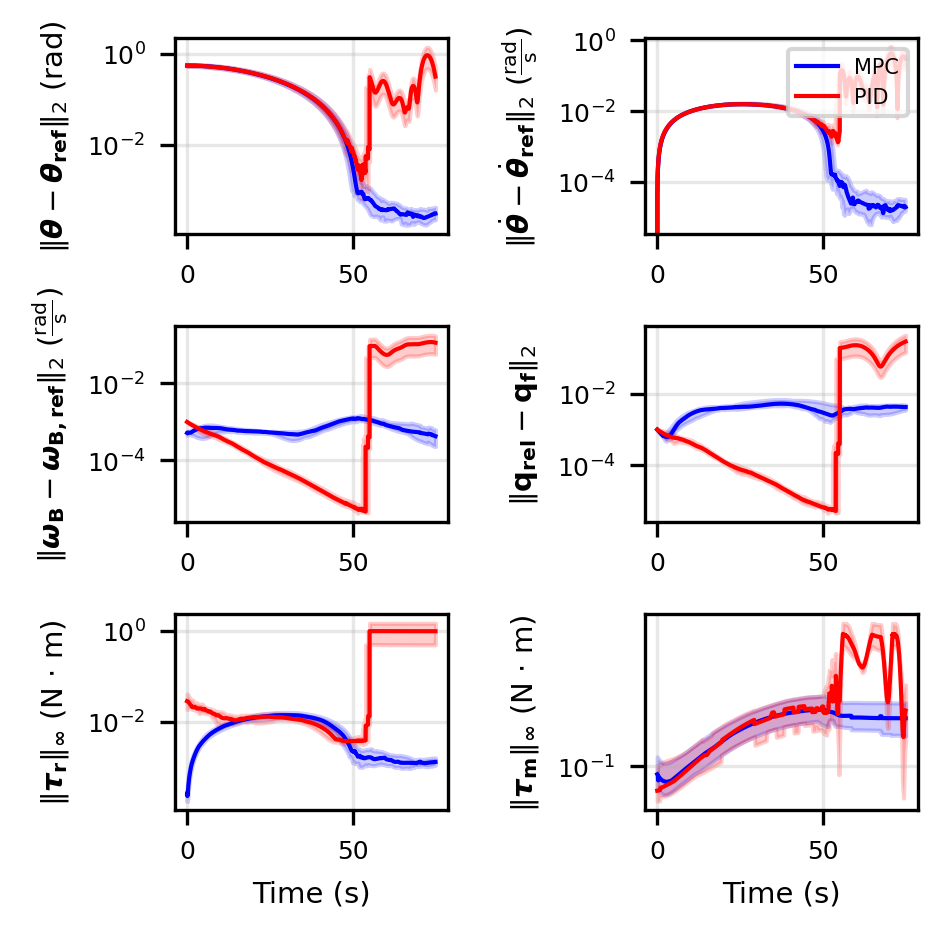}
        \label{fig:mc_case_a_phase_b}
    \end{subfigure}
    \caption{Case study A results. Phase A (first column): servicer satellite angular velocity two-norm error (top), relative orientation quaternion two-norm error (middle), and RW torque infinity norm (bottom) versus operation horizon $\mc{T}_A$. Phase B (second column): joint angle two-norm error (top),  servicer satellite angular velocity two-norm error (middle), and RW torque infinity norm (bottom) versus operation horizon $\mc{T}_B$. Phase B (third column): joint velocity two-norm error (top), relative orientation quaternion two-norm error (middle), joint torque infinity norm (bottom) versus versus operation horizon $\mc{T}_B$.}
    \label{fig:case_a_mc}
\end{figure}
The results of case study A listed in Table~\ref{tab:all_case_studies} and visualized in Figure~\ref{fig:case_a_mc} show that the MPC controller provides significant advantages for both phases while maintaining strict safety margins. In Table~\ref{tab:all_case_studies}, we notice that the MPC controller had $0\%$ constraint violation occurrences across all successful runs, whereas the PID controller had $0.4\%$ constraint violation occurrences across all successful runs. As shown in Figure~\ref{fig:case_a_mc} for phase A results, the MPC controller was able to complete spin synchronization faster than the PID controller while expending lesser control effort. Furthermore, the MPC controller was able to then achieve zero-impulse contact with the target satellite while ensuring spin synchronization was maintained, while PID completes the coordinated maneuver with greater tracking error than MPC according to Figure~\ref{fig:case_a_mc} for phase b results. This is further indicated by the results in Table~\ref{tab:all_case_studies}, where MPC has $2.56$ times lower relative orientation quaternion error and $7.69$ times lower angular velocity error than PID, as well as $17\%$ lower end effector position error and $25\%$ lower end effector velocity error compared to PID. Regarding mission reliability, according to Table~\ref{tab:all_case_studies}, the MPC controller achieved an $86\%$ success rate in case study A, with the $14\%$ of failures resulting from OCP solver failures; in contrast, the PID controller achieved only a $64\%$ success rate in case study A, with $36\%$ of failures attributed to state-divergence issues. The only advantage that PID has over MPC is that PID required a lesser computation time per control cycle on average, where MPC required $1.0$ second per control cycle and PID required $0.018$ second per control cycle according to Table~\ref{tab:all_case_studies}. Thus, further work must be done to ensure MPC has comparable average computation times to PID to ensure computational tractability. Overall, case study A's results indicate that compared to PID,  MPC is better at achieving zero constraint violations, lower errors in relative orientation quaternion, angular velocity, and end effector position and velocity according to Table~\ref{tab:all_case_studies}. 

\begin{figure}[ht!]
    \centering
    \begin{subfigure}[t]{0.48\linewidth}
        \centering
        \includegraphics[width=\linewidth]{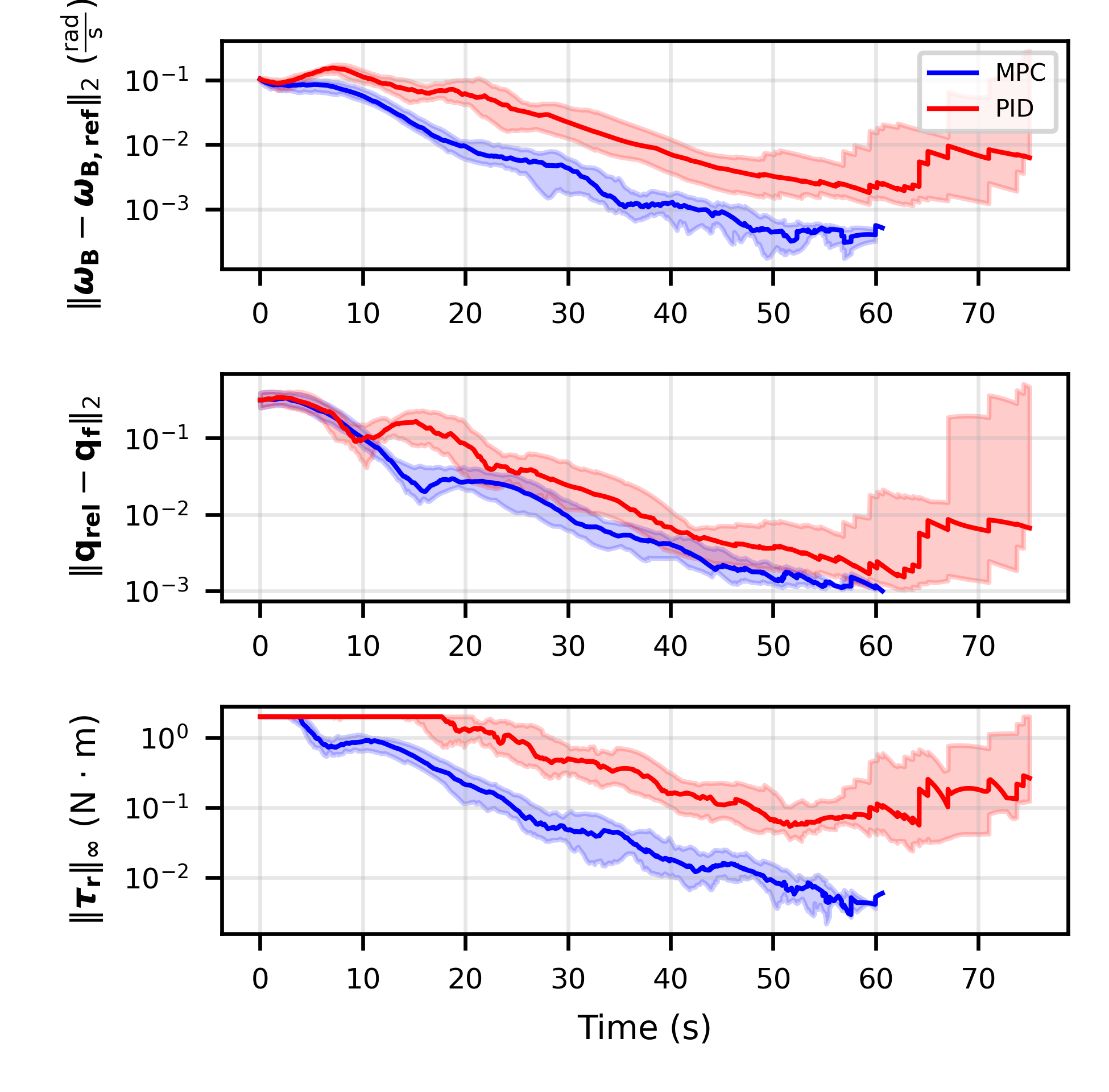}
        \label{fig:mc_case_b_phase_a}
    \end{subfigure}
    \hfill
    \begin{subfigure}[t]{0.48\linewidth}
        \centering
        \includegraphics[width=\linewidth]{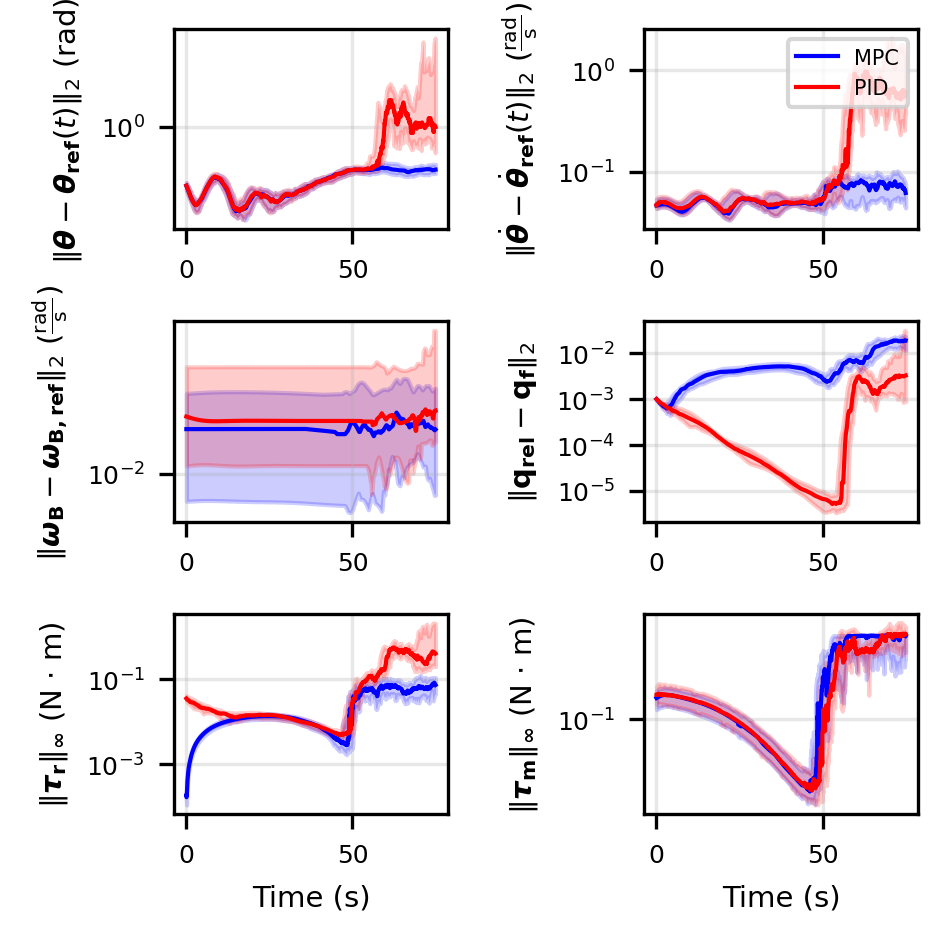}
        \label{fig:mc_case_b_phase_b}
    \end{subfigure}
    \caption{Case study B results. Phase A (first column): servicer satellite angular velocity two-norm error (top), relative orientation quaternion two-norm error (middle), RW torque infinity norm (bottom) versus operation horizon $\mc{T}_A$. Phase B (second column): joint angle two-norm error (top),  servicer satellite angular velocity two-norm error (middle), RW torque infinity norm (bottom) vs operation horizon $\mc{T}_B$. Phase B (third column): joint velocity two-norm error (top), relative orientation quaternion two-norm error (middle), joint torque infinity norm (bottom) versus operation horizon $\mc{T}_B$.}
    \label{fig:case_b_mc}
\end{figure}
The results of case study B listed in Table~\ref{tab:all_case_studies} and visualized in Figure~\ref{fig:case_b_mc} show that the MPC controller provides significant advantages for both mission phases while maintaining strict safety margins, especially when tracking a time-varying trajectory in phase B. We notice that in the results for phase B according to Figure~\ref{fig:case_b_mc}, the MPC controller does a better job at matching the reference trajectory compared to PID, since the trajectory is initially followed by both controllers, but PID completes the phase with greater joint angle and joint velocity error than the MPC controller. These results are further indicated by Table~\ref{tab:all_case_studies}, where achieved $7.89\%$ lower end effector position error and two-times lower end effector velocity error compared to PID. Furthermore, as shown in Figure~\ref{fig:case_b_mc}, both MPC and PID controllers were able to maintain spin synchronization while achieving zero-impulse contact, with MPC achieving $5.45$-times lower angular velocity error and $4.81$-times lower relative orientation quaternion error according to Table~\ref{tab:all_case_studies}. As listed in Table~\ref{tab:all_case_studies}, the MPC controller had $0\%$ of constraint violation occurrences across all successful runs, while PID had $8.03\%$ constraint violation occurrences across all successful runs. Regarding mission reliability, according to Table~\ref{tab:all_case_studies}, the MPC controller achieved an $92\%$ success rate, with the $8\%$ of failures resulting from OCP solver failures; in contrast, the PID controller achieved only a $80\%$ success rate, with $20\%$ of failures attributed to state-divergence issues. The only advantage that PID has over MPC is that PID required a lesser computation time per control cycle on average, where MPC required $0.72$ second per control cycle and PID required $0.019$ second per control cycle according to Table~\ref{tab:all_case_studies}. Thus, further work must be done to ensure MPC has comparable average computation times to PID to ensure computational tractability when tracking time-varying trajectories. Overall, case study B's results indicate that compared to PID, MPC is better at tracking a time-varying trajectories  while achieving lower relative orientation quaternion error, angular velocity error, end effector position and velocity error, and zero constraint violations according to Table~\ref{tab:all_case_studies}.

\begin{figure}[ht!]
    \centering
    \begin{subfigure}[t]{0.48\linewidth}
        \centering
        \includegraphics[width=\linewidth]{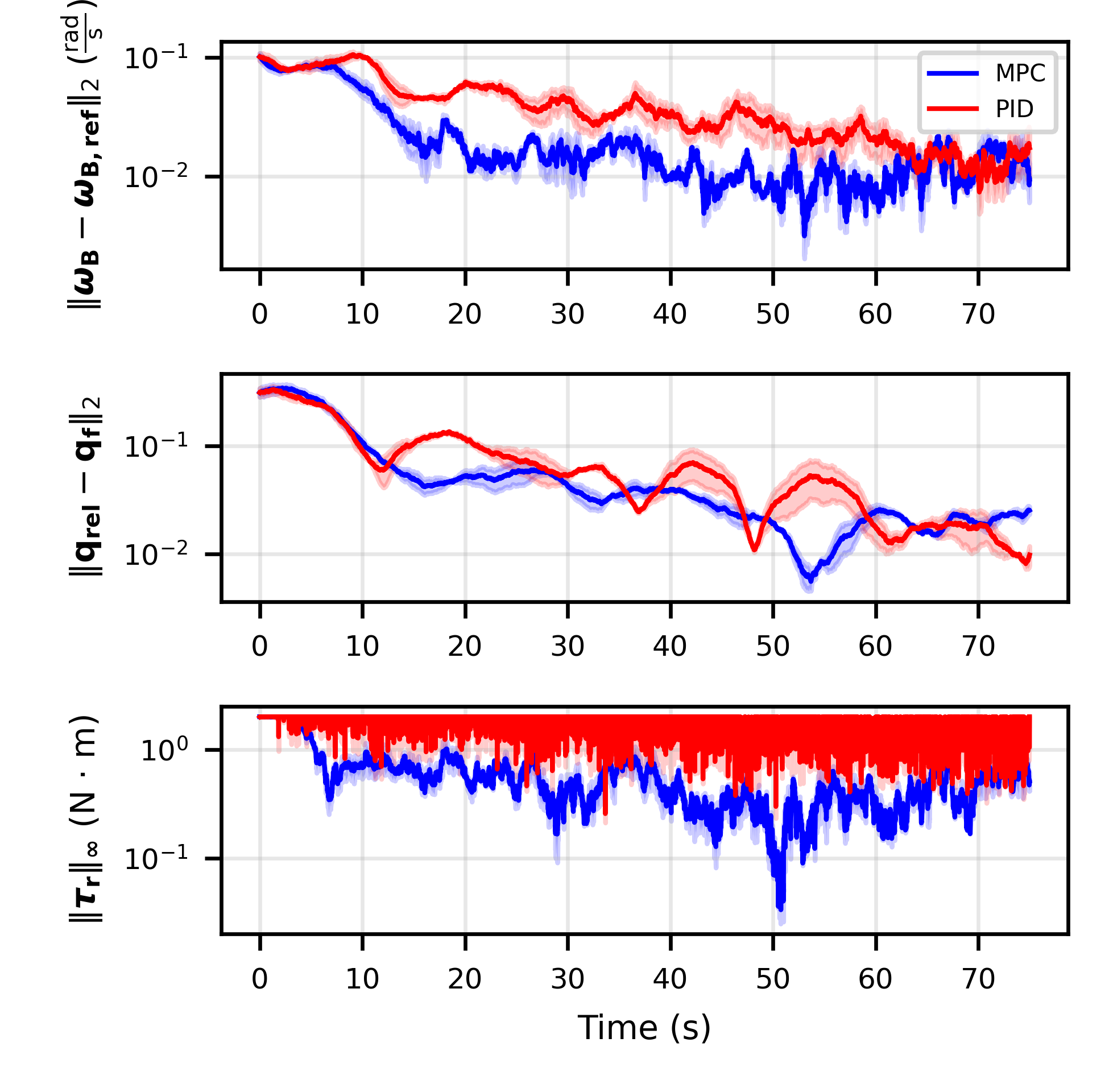}
        \label{fig:mc_case_c_phase_a}
    \end{subfigure}
    \hfill
    \begin{subfigure}[t]{0.48\linewidth}
        \centering
        \includegraphics[width=\linewidth]{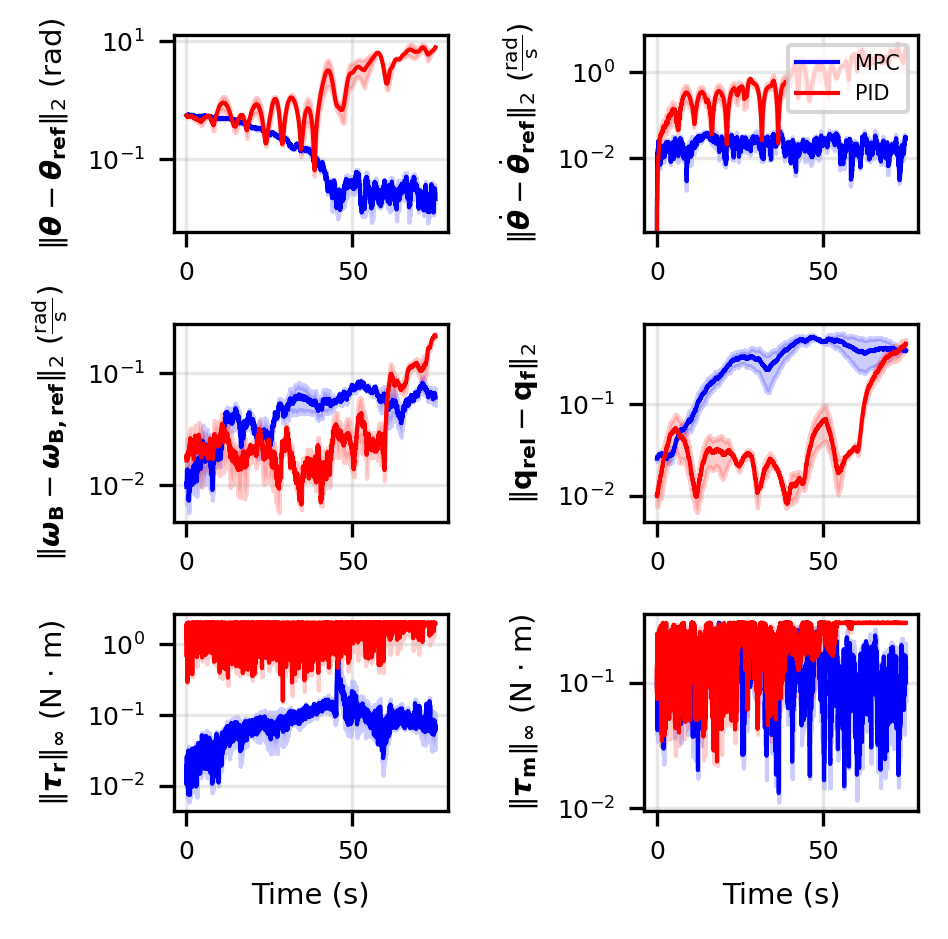}
        \label{fig:mc_case_c_phase_b}
    \end{subfigure}
    \caption{Case study C results.  Phase A (first column): servicer satellite angular velocity two-norm error (top), relative orientation quaternion two-norm error (middle), RW torque infinity norm (bottom) versus operation horizon $\mc{T}_A$. Phase B (second column): joint angle two-norm error (top), servicer satellite angular velocity two-norm error (middle),  RW torque infinity norm (bottom) vs operation horizon $\mc{T}_B$. Phase B (third column): joint velocity two-norm error (top),  relative orientation quaternion two-norm error (middle), joint torque infinity norm (bottom) versus operation horizon $\mc{T}_B$.}
    \label{fig:case_c_mc}
\end{figure}
The results of case study C listed in Table~\ref{tab:all_case_studies} and visualized in Figure~\ref{fig:case_c_mc} show that MPC provides substantially greater robustness to noisy conditions. MPC is able to operate under noisy conditions while having zero constraint violation occurrences, while PID operates under noisy conditions with $26.41\%$ constraint violation occurrences. Regarding the mission reliability under noisy conditions, according to Table~\ref{tab:all_case_studies} for case study C, achieved a $60\%$ success rate, with the $40\%$ of failures resulting from OCP solver failures. In contrast, the PID controller achieved only a $10\%$ success rate, with the $90\%$ of failures attributed to state-divergence issues. According to Table~\ref{tab:all_case_studies} and Figure~\ref{fig:case_c_mc}, the phase A results show that the MPC and PID controllers were able to achieve spin synchronization with comparable angular velocity errors and relative orientation quaternion errors. However, under noisy conditions, the PID controller expended more torque than the MPC, which is also a recurring pattern in the phase B results according to Figure~\ref{fig:case_c_mc}. Furthermore, during phase B according to Figure~\ref{fig:case_c_mc}, the MPC controller was able to achieve lower joint angle, joint velocity, and angular velocity tracking error, which is further supported in Table~\ref{tab:all_case_studies}, where MPC achieved $1.07$-times lesser angular velocity error, $3.17$-times lesser end effector position error, and $20.5$-times lesser end-effector velocity error compared to PID, indicating that the MPC controller was able to complete both mission phases under noisy conditions. However, an area of concern is that the PID controller achieved lesser relative orientation quaternion error compared to the MPC controller, with PID achieving $3.01$-times lesser error than MPC according to Table~\ref{tab:all_case_studies}. This is further supported by Figure~\ref{fig:case_c_mc}, where the MPC controller has greater relative orientation quaternion error compared to PID throughout the entire phase B simulation, indicating that  future work should aim to improve on the relative orientation quaternion error for MPC under noisy conditions. Another advantage that PID has over MPC is that PID required a lesser computation time per control cycle on average, where MPC required $0.49$ second per control cycle and PID required $0.018$ second per control cycle according to Table~\ref{tab:all_case_studies}. Thus, further work must also be done to ensure MPC has comparable average computation times to PID to ensure computational tractability under noisy conditions. Overall, the results in case study C indicate that MPC outperforms PID under noise by showing lower angular velocity error and lower end effector position and velocity error, as well as achieving safe operations with zero constraint violations according to Table~\ref{tab:all_case_studies}.

\section{Conclusion}
This project developed and implemented a nonlinear MPC framework for achieving safe, constraint-compliant zero-impulse contact with a free-spinning target satellite. Through comprehensive MC evaluation across three case studies, we demonstrated that the MPC controller fundamentally outperforms prior control approaches for OOS operations. The key differentiator is the MPC controller's explicit incorporation of momentum-coupled dynamics and actuation constraints directly into the optimization problem, providing the predictive capability necessary to maintain operational safety margins that prior control approaches cannot guarantee. While our MPC framework achieves superior performance, it requires solving a large-scale optimization problem per phase, resulting in greater computational demands than prior approaches. This increased complexity poses challenges for on-board implementation with limited computational resources. Future work should investigate decomposition strategies to enhance computational tractability and experimental hardware validation to assess real-time feasibility.


\bibliography{sample}

\appendix
\section{Inertia matrix blocks of the servicer satellite}
\label{APP:inertia_deriv}

We collect here the explicit expressions of the blocks of the inertia matrix $H$ introduced in equation~\eqref{eq:kinetic_energy}, with block structure shown in equation~\eqref{eq:First_dynamics}. We first define the skew-symmetric operator as
\begin{equation}
  [\bm r \times] =
  \begin{bmatrix}
    0 & -r_z & r_y \\
    r_z & 0  & -r_x \\
    -r_y & r_x & 0
  \end{bmatrix},
  \qquad
  D(\bm r) = [\bm r \times]^\top [\bm r \times].
\end{equation}
We index the rigid bodies of the servicer satellite by $i \in [0,6]$, where $i=0$ denotes the moment-generation base, $i \in \mathcal{L} := [1,3]$ the three manipulator links, and $i \in \mathcal{R} := [4,6]$ the three reaction wheels of the RW cluster.

For each body $i$, let  $m_i \in \reals_+$ be its mass, $I_i \in \reals^{3\times 3}$ be its inertia matrix in the moment-generation base frame $B$, $\bm{r}_i \in \reals^3$ be the position of its CoM with respect to $B$.
\begin{equation}
     H_V = \Big( \sum_{i=0}^{6} m_i \Big) I_{3\times 3} \in \reals^{3\times 3}, \quad
     H_{\textbf{V}{\bm{\Omega}}} = - \sum_{i=0}^{6} m_i [\bm{r}_i \times] \in \reals^{3\times 3}, \quad
     H_{\textbf{V}{\bm{\theta}}} = \sum_{i \in \mathcal{L}} m_i J_{L_i}(\bm{\theta}) \in \reals^{3\times 3}, \quad
     H_{\textbf{V}{\bm{\phi}}} = 0_{3\times 3} \in \reals^{3\times 3},
\end{equation}
\begin{equation}
    H_{\bm{\Omega}} = \sum_{i=0}^{6} \big( I_i + m_i D(\bm{r}_i) \big) \in \reals^{3\times 3}, \,
    H_{{\bm{\Omega}}{\bm{\theta}}} = \sum_{i \in \mathcal{L}} \Big( I_i J_{A_i}(\bm{\theta}) + m_i [\bm{r}_i \times] J_{L_i}(\bm{\theta}) \Big) \in \reals^{3\times 3}, \,
    H_{{\bm{\Omega}}\bm{\phi}} = \sum_{k \in \mathcal{R}} I_k J_{A_k}(\bm{\bm{\phi}}) \in \reals^{3\times 3},
\end{equation} where the value of $H_{\textbf{V}\bm{\phi}}$ follows from the assumption that the RW cluster is purely rotational and does not contribute to the translational kinetic energy of the servicer satellite.
The manipulator inertia matrix (for a fixed moment-generation base frame $B$) can be written as
\begin{equation}
  H_{\bm{\theta}}(\bm{\theta})
  = \sum_{i \in \mathcal{L}} \Big(
    J_{L_i}(\bm{\theta})^\top m_i J_{L_i}(\bm{\theta})
    + J_{A_i}(\bm{\theta})^\top I_i J_{A_i}(\bm{\theta})
  \Big) \in \reals^{3\times 3}.
\end{equation}
We assume that there is no kinetic-energy coupling between the manipulator arm's joint rates and the RW cluster's rates, thus $H_{{\bm{\theta}}\bm{\phi}} = 0_{3\times 3}$.
Finally, the RW cluster block (for a fixed moment-generation base frame $B$) reflects its rotational contribution to the total kinetic energy of the servicer satellite as
\begin{equation}
  H_{\bm{\phi}}(\bm{\phi})
  = \sum_{k \in \mathcal{R}} J_{A_k}(\bm{\phi})^\top\, I_k\, J_{A_k}(\bm{\phi}) \in \reals^{3\times 3}.
\end{equation}
Because each wheel spins about the principal axis of the moment-generation base, the spin axes are decoupled and thus $H_{\bm{\phi}}$ reduces to a diagonal matrix, with the diagonal elements being each RW's inertia about its spin axes.

Using these blocks, the generalized inertia matrix and nonlinear terms introduced in equation~\eqref{eq:generalized_inertia_M} take the form
\begin{equation}
  \begin{bmatrix}
    M_b & M_{bm} & M_{br} \\
    M_{bm}^\top & M_m & 0_{3\times 3} \\
    M_{br}^\top & 0_{3\times 3} & M_r
  \end{bmatrix}
  =
  \begin{bmatrix}
    H_{{\bm{\Omega}}} - H_{\textbf{V} {\bm{\Omega}}}^\top H_V^{-1} H_{\textbf{V} {\bm{\Omega}}} &
    H_{{\bm{\Omega}} {\bm{\theta}}} - H_{\textbf{V} {\bm{\Omega}}}^\top H_V^{-1} H_{\textbf{V} {\bm{\theta}}} &
    H_{{\bm{\Omega}} r} \\
    H_{{\bm{\Omega}} {\bm{\theta}}}^\top - H_{\textbf{V} {\bm{\theta}}}^\top H_V^{-1} H_{\textbf{V} {\bm{\theta}}} &
    H_{{\bm{\theta}}} - H_{\textbf{V} {\bm{\theta}}}^\top H_V^{-1} H_{\textbf{V} {\bm{\theta}}} &
    0_{3 \times 3} \\
    H_{{\bm{\Omega}} r}^\top & 0_{3 \times 3} & H_r
  \end{bmatrix}, \quad \begin{bmatrix}
    \bm{c}_b \\ \bm{c}_m \\ \bm{c}_r
  \end{bmatrix}
  =
  \begin{bmatrix}
    \bm{\bar{c}}_b - H_{\textbf{V} \bm{\omega}}^\top H_V^{-1} \bm{c}_V \\
    \bm{\bar{c}}_m - H_{\textbf{V} {\bm{\theta}}}^\top H_V^{-1} \bm{c}_V \\
    \bm{c}_r
  \end{bmatrix}.
\end{equation}

\section{MPC solver summary}
\label{APP:_MPC_solver}
The \texttt{acados} solver efficiently solves complex nonlinear optimization problems. To handle the system's nonlinear dynamics, it applies the \textbf{Gauss-Newton method} to approximate second-order derivatives, treating the objective function as a nonlinear least-squares problem. To ensure convergence of the nonlinear solver, it uses \textbf{globalization techniques}, which adjusts the way the solver tries to find solutions, by modifying the step size or search direction, if the initial guesses lead to divergence. For numerical integration, it utilizes the \textbf{implicit Runge-Kutta method (IRK)} to find discretized numerical solutions to continuous differential equations. IRK methods exhibit superior stability properties compared to explicit integrators, particularly for the stiff differential equations arising from the inertia matrix inversions in our coupled dynamics, enabling larger timesteps without sacrificing solution accuracy.

For solution techniques, \texttt{acados} uses SQP, an iterative approach that breaks down the constrained nonlinear optimization problem into smaller, more manageable \textbf{quadratic programming} (QP) subproblems. Key parameters characterize the process, including limits on the number of iterations for the nonlinear solver and the tolerance defining how closely the solution must meet the constraints.
It also uses the \textbf{interior-point method} (IPM) which is similar to the SQP method, but adds a barrier function to the objective and advances through the interior of the feasible region, avoiding the boundary, until the optimal solution is found. This method offers a polynomial runtime and is suitable for solving linear and nonlinear convex optimization problems.
\subsection{Solver configuration}
Table~\ref{tab:solver_options} summarizes the main solver configurations used in the nonlinear MPC formulation. Specifically, we used the Gauss–Newton Hessian approximation, a merit-function backtracking line-search globalization strategy, an SQP-based nonlinear solver with IRK integration, and a high-performance IPM (HPIPM) QP solver. The solver tolerances are relatively strict, and the maximum numbers of NLP/QP iterations are chosen sufficiently large to ensure convergence of the optimization problem for all the following considered scenarios.
\begin{table}[ht!]
\centering
\resizebox{0.8\textwidth}{!}{
\begin{tabular}{@{}|l|l|c|@{}}
\hline
\multicolumn{3}{|c|}{\textbf{Hessian Approximation}} \\ \hline
\texttt{hessian\_approx} & Type of Hessian approximation & \texttt{GAUSS\_NEWTON} \\ \hline
\texttt{globalization} & Turns on globalization & \texttt{MERIT\_BACKTRACKING} \\ \hline
\texttt{regularize\_method} & Regularization method used & \texttt{MIRROR} \\ \hline

\multicolumn{3}{|c|}{\textbf{Nonlinear Solver}} \\ \hline
\texttt{nlp\_solver\_type} & Type of nonlinear solver & \texttt{SQP} \\ \hline
\texttt{nlp\_solver\_max\_iter} & Maximum number of iterations for the solver & 2000 \\ \hline
\texttt{nlp\_solver\_tol\_eq} & Tolerance for equality constraints & \( 10^{-5} \) \\ \hline

\multicolumn{3}{|c|}{\textbf{Integrator Type}} \\ \hline
\texttt{integrator\_type} & Type of integrator used & \texttt{IRK} \\ \hline
\texttt{sim\_method\_num\_stages} & Number of stages in the integrator & 4 \\ \hline
\texttt{sim\_method\_num\_steps} & Number of steps in the integrator & 2 \\ \hline

\multicolumn{3}{|c|}{\textbf{QP Solver}} \\ \hline
\texttt{qp\_solver} & Type of quadratic program solver & \texttt{PARTIAL\_CONDENSING\_HPIPM} \\ \hline
\texttt{qp\_solver\_cond\_N} & QP solver conditioning horizon & \texttt{N\_horizon} \\ \hline
\texttt{qp\_solver\_max\_iter} & Maximum number of iterations for the solver & 2000 \\ \hline
\texttt{qp\_solver\_tol\_eq} & Tolerance for equality constraints & \( 10^{-5} \) \\ \hline

\multicolumn{3}{|c|}{\textbf{Solver Tolerance}} \\ \hline
\texttt{tol} & Tolerance for the solver & \( 10^{-5} \) \\ \hline

\multicolumn{3}{|c|}{\textbf{Shooting Intervals and Horizon}} \\ \hline
\texttt{N\_horizon} & Number of shooting intervals & 70 \\ \hline
\texttt{tf} & Prediction horizon time & 0.70 \\ \hline
\end{tabular}
}
\caption{Solver Options for \texttt{acados}}
\label{tab:solver_options}
\end{table}
\section{PID tuning description}
\label{APP:PID_tuning_details}
For the manipulation module, we treat the arm as fixed-base during PID tuning and use the simplified control law
\begin{equation} \label{eq:Arm_torques}
    \bm{\tau_m}(t) = \bm{c_m} - M_m\left(
        \bm{\ddot{\theta}_{\textbf{ref}}}(t) + k_d\big(\bm{\dot{\theta}_{\textbf{ref}}}(t) - \bm{\dot{\theta}}(t)\big)
        + k_p\big(\bm{\theta_{\textbf{ref}}}(t) - \bm{\theta}(t)\big)
        + k_i \int_0^T \big(\bm{\theta_{\textbf{ref}}}(x) - \bm{\theta}(x)\big)\, dx
    \right),
\end{equation}
which captures the dominant arm dynamics through $M_m$, while coupling effects appear mainly as disturbances that the integral action compensates.

We use the closed-loop Ziegler–Nichols method~\cite{copeland2008design} for tuning. Integral and derivative gains are initially set to zero, and the proportional gain is increased until sustained oscillations occur, yielding the ultimate gain $k_{cr}$ and ultimate period $T_{cr}$. The PID gains are then chosen as
\begin{align}
    k_p = 0.6 k_{cr}, \quad
    k_i = \frac{1.2 k_{cr}}{T_{cr}}, \quad
    k_d = 0.075 k_{cr} T_{cr}.
\end{align}
For the moment-generation base in Phase A, tuning yielded critical gains $k_{q,cr} = 0.6$ and $k_{{\bm{\omega}},cr} = 0.005$ with ultimate period $T_{cr} = 100$\,s. For the manipulator, we obtained $k_{p,cr} = 0.864$ with $T_{cr} = 15.6$\,s. Applying the Ziegler–Nichols formulas produces the final gains
\begin{align}
    k_{\textbf{q}} &= 0.396, \quad k_{\bm{\omega}} = 0.0033, \quad k_{i,{\textbf{q}}} = 0.0396, \quad k_{i,{\bm{\omega}}} = 0.00033, \\
    k_{d,{\textbf{q}}} &= 0.99, \quad k_{d,{\bm{\omega}}} = 0.00825, \\
    k_p &= 0.57024, \quad k_i = 0.097812, \quad k_d = 0.299376.
\end{align}
These values are used in all PID simulations reported in the case studies.
\section{Simulation parameters}
\label{APP:sim_params}
Table~\ref{tab:parameters} summarizes the key nominal parameters used in our simulations. Specifically, we define the sizing terms for the moment-generation base, the manipulation arm, and the RW cluster. The moment-generation base parameters comprise its mass and dimensions (height, length, and width). The manipulation arm parameters include the masses, radii, and lengths of the joints that compose the arm. Lastly, the RW cluster parameters provide information on the mass, radii, height, and distances of the RWs from the center of mass (CoM) of the servicer satellite along the $x,y,z$-axes.
We base the manipulation arm and moment-generation base nominal parameters on the data presented in~\cite{Aghili2024}, and select the RW cluster nominal parameter values freely by referencing and comparing different RW construction sheets.
\begin{table}[ht!]
\centering
\caption{Nominal System Parameters}
\label{tab:parameters}
\setlength{\tabcolsep}{3pt}
\renewcommand{\arraystretch}{1.05}
\small
\begin{tabular}{|l|l|c|}
\hline
\textbf{Parameter} & \textbf{Description} & \textbf{Value} \\ \hline
\multicolumn{3}{|c|}{\textbf{Manipulation Arm Nominal Parameters}} \\ \hline
$m_1, m_2, m_3$       & Masses of manipulation-arm segments              & $1, 3, 2~\mathrm{kg}$ \\
$rad_1, rad_2, rad_3$ & Radii of manipulation-arm joints                 & $0.2, 0.3, 0.4~\mathrm{m}$ \\
$L_1, L_2, L_3$       & Lengths of manipulation-arm segments             & $0.2, 0.8, 0.5~\mathrm{m}$ \\ \hline
\multicolumn{3}{|c|}{\textbf{Moment-Generation Base Nominal Parameters}} \\ \hline
$m_b$                 & Mass of the satellite base                       & $150.0~\mathrm{kg}$ \\
$h_b, l_b, w_b$       & Dimensions: height, length, width                & $1.9, 2.45, 1.41~\mathrm{m}$ \\ \hline
\multicolumn{3}{|c|}{\textbf{Reaction-Wheel (RW) Cluster Nominal Parameters}} \\ \hline
$m_{\text{RW}}$       & Mass of each RW                                  & $5.0~\mathrm{kg}$ \\
$r_{\text{RW1}}, r_{\text{RW2}}$
                      & Radii (small, large)                             & $0.337/3, 0.337/2~\mathrm{m}$ \\
$h_{\text{RW}}$       & Height of each RW                                & $0.1~\mathrm{m}$ \\
$r_{\text{RW}_x}, r_{\text{RW}_y}, r_{\text{RW}_z}$
                      & Distances from CoM in $x,y,z$ directions         & $w_b/8, l_b/8, h_b/8$ \\ \hline
\end{tabular}
\end{table}
\section{MPC solver parameters}
\label{APP:solve_params}
Tables~\ref{OCP_A} and~\ref{OCP_B} define the state and input constraints, cost matrices and the initial conditions used in Section~\ref{sec:sim}. The $Q$ and $R$ values depend on the specific phase and are provided in the simulation section. 
\begin{table}[ht!]
\centering
\small
\begin{tabular}{|>{\raggedright\arraybackslash}m{4cm}|>{\raggedright\arraybackslash}m{11cm}|}
\hline
\textbf{Variable} & \textbf{Values} \\
\hline
Initial Condition, \( \bm{x_0} \) & \( \begin{bmatrix}
    0.1 & 0.0 & 0.2 &
    0.1 & 0.1 & 0.1 & 1.0
\end{bmatrix}_{7 \times 1} \) \\
\hline
State Constraints, \( \bm{x_{\textbf{min}}} \) & \( \begin{bmatrix}
    -0.5 & -0.5 & -0.5 &
    -0.9 & -0.9 & -0.9 & -1.0
\end{bmatrix}_{7 \times 1} \) \\
\hline
State Constraints, \( \bm{x_{\textbf{max}}} \) & \( \begin{bmatrix}
    0.5 & 0.5 & 0.5 &
    0.9 & 0.9 & 0.9 & 1.0
\end{bmatrix}_{7 \times 1} \) \\
\hline
Control Input Constraints, \( \bm{\tau_{\textbf{min}}} \) & \( \begin{bmatrix}
    -2 & -2 & -2
\end{bmatrix}_{3 \times 1} \) \\
\hline
Control Input Constraints, \( \bm{\tau_{\textbf{max}}} \) & \( \begin{bmatrix}
    2 & 2 & 2
\end{bmatrix}_{3 \times 1} \) \\
\hline
State Cost Matrix, \( Q \) & \( 400 \cdot \text{block\_diag}(
7, 7, 9, 9, 9, 12, 15) \) \\
\hline
Control Cost Matrix, \( R \) & \( 2 \cdot \text{block\_diag}(0.8, 0.4, 0.6) \) \\
\hline
Weight Matrix, \( W \) & \( \text{block\_diag}(Q, R) \) \\
\hline
Terminal Cost Matrix, \( W_e \) & \( Q \) \\
\hline
\end{tabular}
\caption{OCP – phase A cost matrices, initial conditions, state and control constraints.}
\label{OCP_A}
\end{table}
\hfill
\begin{table}[ht!]
\centering
\small
\begin{tabular}{|>{\raggedright\arraybackslash}m{4cm}|>{\raggedright\arraybackslash}m{11cm}|}
\hline
\textbf{Variable} & \textbf{Values} \\
\hline
Initial Condition, \( \bm{x_0} \) & \( 
\Bigl[\,0.05 \quad 0.4 \quad 0.05 \quad 0.0 \quad 0.0 \quad 0.2 \quad 0.0 \quad 0.0 \quad 0.0 \quad 0.0 \quad 0.0 \quad 0.0 \quad 1.0\,\Bigr]_{13 \times 1} \) \\
\hline
State Constraints, \( \bm{x_{\textbf{min}}} \) & \( 
\Bigl[\, -0.8\,-0.8\,-0.8\,-0.5\,-0.5\,-0.5\,-0.8\,-0.8\,-0.8\,-0.9\,-0.9\,-0.9\,-1.0 \,\Bigr]_{13 \times 1} \) \\
\hline
State Constraints, \( \bm{x_{\textbf{max}}} \) & \( 
\Bigl[\,0.8 \quad 0.8 \quad 0.8 \quad 0.5 \quad 0.5 \quad 0.5 \quad 0.8 \quad 0.8 \quad 0.8 \quad 0.9 \quad 0.9 \quad 0.9 \quad 1.0\,\Bigr]_{13 \times 1} \) \\
\hline
Control Input Constraints, \( \bm{\tau_{\textbf{min}}} \) & \( \begin{bmatrix}
    -2 & -2 & -2 & -0.3 & -0.3 & -0.3
\end{bmatrix}_{6 \times 1} \) \\
\hline
Control Input Constraints, \( \bm{\tau_{\textbf{max}}} \) & \( \begin{bmatrix}
    2 & 2 & 2 & 0.3 & 0.3 & 0.3
\end{bmatrix}_{6 \times 1} \) \\
\hline
State Cost Matrix, \( Q \) & \( 400 \cdot \text{block\_diag}(
20, 20, 25, 21, 21, 27, 15, 15, 15, 27, 27, 27, 32) \) \\
\hline
Control Cost Matrix, \( R \) & \( 2 \cdot \text{block\_diag}(100, 100, 100, 20, 20, 20) \) \\
\hline
Weight Matrix, \( W \) & \( \text{block\_diag}(Q, R) \) \\
\hline
Terminal Cost Matrix, \( W_e \) & \( Q \) \\
\hline
\end{tabular}
\caption{OCP – phase B cost matrices, initial conditions, state and control constraints.}
\label{OCP_B}
\end{table}

The weight matrices $Q$ and $R$ are tuned to specific phase's control priorities. In phase A, spin synchronization requires precise alignment of the servicer satellite's angular velocity $\bm{\omega_B}(t)$ and relative quaternion orientation $\bm{q_{\textbf{rel}}}(t)$ to the target satellite. Thus we assign high weights to errors in angular velocity and relative quaternion orientation.
Moment-generation module torque costs remain modest to permit aggressive maneuvering during spin synchronization.

In phase B, accurate joint tracking becomes critical, as joint angle $\bm{\theta}(t)$ and joint velocity $\bm{\dot{\theta}}(t)$ errors propagate through the manipulation arm to the end effector. We assign moderate weights to penalize slight deviations in joint angles and joint velocities while maintaining spin synchronization. Moment-generation module torque costs are increased relative to manipulation module torque costs to prevent disturbances from the moment-generation module during manipulation-arm motion.
\end{document}